\documentclass[english,showkeys,notitlepage]{revtex4-1}
\usepackage[T1]{fontenc}
\usepackage[latin9]{inputenc}
\usepackage{babel}
\usepackage{amsmath}
\usepackage{amssymb}
\usepackage{graphicx}
\usepackage{esint}
\usepackage[unicode=true,pdfusetitle,
 bookmarks=true,bookmarksnumbered=false,bookmarksopen=false,
 breaklinks=false,pdfborder={0 0 1},backref=false,colorlinks=false]
 {hyperref}

\makeatletter

\usepackage{babel}
\makeatother

\begin{document}

\title{Electroproduction of $D$- and $B$-mesons in high-multiplicity $ep$
collisions}

\author{Marat Siddikov and Iván Schmidt}

\affiliation{Departamento de Física, Universidad Técnica Federico Santa María,~~\\
 y Centro Científico - Tecnológico de Valparaíso, Casilla 110-V, Valparaíso,
Chile}
\begin{abstract}
In this paper we study the electroproduction of open heavy flavor
$D$- and $B$-mesons in the kinematics of future $ep$ colliders,
such as the Electron Ion Collider (EIC), the Large Hadron electron Collider (LHeC)
and the Future Circular Collider (FCC-he). We study in detail the dependence
of the cross-sections on multiplicity of co-produced hadrons, in view
of its possible sensitivity to contributions from multipomeron contributions,
and discuss different observables which might be used for its study.
According to our theoretical expectations, in $ep$ collisions the
multipomeron contributions are small in the EIC kinematics, although they might
be sizable at LHeC and FCC-he. We also provide theoretical predictions
for the production cross-sections of heavy mesons in the kinematics
of all the above-mentioned $ep$ colliders.

\end{abstract}
\maketitle

\section{Introduction}

Due to the high luminosity of the forthcoming LHC upgrade (HL-LHC) and
future electron-proton colliders, many rare processes recently got
renewed theoretical interest. One of the directions which might benefit
from the outstanding luminosity is the production of different hadrons
in high-multiplicity events. The development of theoretical framework
for the study of such events was initiated more than forty years
ago in~\cite{Abramovsky:1973fm,Capella:1976ef,Bertocchi:1976bq,Shabelski:1977iv,Nikolaev:2006mx,Kaidalov:1982xe}.
However, for a long time the experimental study of such processes was limited
by the insufficient luminosity of existing high-energy experiments (see
however the discussion in~~\cite{Bartels:1996hw,Bartels:2005wa,Kovchegov:1999yj,Kovchegov:2000hz,Kovchegov:2012mbw,Sjostrand:2004pf}
related to HERA). At RHIC and LHC, thanks to the very large luminosity,
the multiplicity dependence of hadroproduction processes has been
studied in great detail, and various elaborate observables have been
measured experimentally, extending our understanding of the mechanisms
of these processes. For example, the experimental study of yields
of light charged hadrons co-produced together with heavier mesons~\cite{Adam:2015ota,Trzeciak:2015fgz,Ma:2016djk,PSIMULT,Khatun:2019slm,Alice:2012Mult}
revealed that the multiplicity dependence is faster than in the absence
of heavy mesons, and, as was suggested in~\cite{Siddikov:2019xvf,Levin:2018qxa,Siddikov:2020pjh,Siddikov:2020lnq,Schmidt:2020fgn},
might be explained by contributions of higher twist multipomeron mechanisms.
This finding is important, because it gives possibility to understand
better the onset of saturation in high-energy collisions. 

It is expected that the future Electron Ion Collider (EIC)~\cite{Accardi:2012qut,AbdulKhalek:2021gbh},
the Large Hadron electron Collider (LHeC)~\cite{AbelleiraFernandez:2012cc}
and the Future Circular Collider (FCC-he)~\cite{Mangano:2017tke,Agostini:2020fmq,Abada:2019lih}
also will have very large luminosities, which will make possible a study
of physics at the intensity frontier in electroproduction processes. The
measurement of the multiplicity dependencies at these new colliders
might be used for better understanding of the underlying microscopic
mechanisms of different \emph{electro}production processes. In what
follows we will focus on the production of heavy flavor $D$- and $B$-mesons,
as well as non-prompt $J/\psi$ mesons. These states might be described
approximately in the heavy quark mass limit~\cite{Korner:1991kf,Neubert:1993mb},
and for this reason have been used since the early days of QCD as
a probe for testing the predictions of perturbative Quantum Chromodynamics
(QCD)~(see \emph{e.g}. ~\cite{Bodwin:1994jh,Maltoni:1997pt,Binnewies:1998vm,Kniehl:1999vf,Brambilla:2008zg,Feng:2015cba,Brambilla:2010cs,Ma:2018bax,Goncalves:2017chx}
for an overview).~ In what follows we will focus on the kinematics
of photoproduction, where most of the heavy mesons are produced from
quasi-real photons with virtuality $Q^{2}\approx0$. In this kinematics
the typical values of Bjorken $x_{B}$ are small, $x_{B}\ll1$, and
the gluon densities significantly exceed the sea quark contributions.
In the proton rest frame the interaction might be viewed as a scattering
of the color dipole, formed from the photon, in the proton gluonic field.
The appropriate description of such process is the color dipole framework
(also known as CGC/Sat)~\cite{GLR,McLerran:1993ni,McLerran:1993ka,McLerran:1994vd,MUQI,MV,gbw01:1,Kopeliovich:2002yv,Kopeliovich:2001ee}.
This approach has been successfully applied to the phenomenological description
of both hadron-hadron and lepton-hadron collisions~\cite{Kovchegov:1999yj,Kovchegov:2006vj,Balitsky:2008zza,Kovchegov:2012mbw,Balitsky:2001re,Cougoulic:2019aja,Aidala:2020mzt,Ma:2014mri},
and allows a straightforward extension for the description of high-multiplicity
events~\cite{KOLEB,KLN,DKLN,Kharzeev:2000ph,Kovchegov:2000hz,LERE,Lappi:2011gu,Ma:2018bax}.
The color dipole approach is not valid for larger values of $x_{B}\gtrsim0.1$,
due to possible contributions of intrinsic quarks (\emph{e.g}. intrinsic
charm). For  this reason in what follows we will  consider only the
variables which do not get significant contributions from that region.
We also will analyze explicitly the role of the multipomeron mechanisms,
which are usually omitted as higher twist effects. Since such contributions
have more pronounced dependence on multiplicity, their presence could
be straightforwardly deduced from experimental data on multiplicity
dependence.  

The paper is structured as follows. In Section~\ref{sec:Evaluation}
we discuss the framework used for the open-heavy meson production evaluation,
taking into account  the contributions of the single-
and double-pomeron  mechanisms,  compare the theoretical expectations
with experimental data and make predictions for the kinematics of the
future electron-proton colliders. In Section~\ref{sec:Numer} we
suggest observables which might help to measure the multiplicity dependence,
and make theoretical predictions for them  in the dipole framework. Finally,
in Section~\ref{sec:Conclusions} we draw conclusions.

\section{Production of open heavy flavor mesons}

\label{sec:Evaluation}The cross-section of open heavy-flavor
meson production via the fragmentation mechanism is given by~\cite{Nikolaev:1995ty,Nikolaev:1994de,Zyla:2020zbs,Binnewies:1998vm,Kniehl:1999vf,Ma:2018bax,Goncalves:2017chx}
\begin{equation}
\frac{d\sigma_{ep\to M+X}}{dx_{B}dy\,d\eta\,d^{2}p_{T}}=\sum_{i}\int_{x_{B}}^{1}\frac{dz}{z^{2}}D_{i}\left(\frac{x_{B}}{z}\right)\,\frac{d\sigma_{ep\to\bar{Q}_{i}Q_{i}+X}}{dx_{B}dy\,d\eta^{*}\,d^{2}p_{T}^{*}}\label{eq:fragConvolution}
\end{equation}
where we use standard DIS notations $Q^{2},\,x_{B},\,y$ for the virtuality
of intermediate photon, Bjorken variable $x_{B}$ and elasticity (fraction
of electron energy which passes to the photon in the proton rest frame);
while $\eta$ and $p_{T}$ are the rapidity and the transverse momentum
of the produced heavy meson. The fragmentation function $D_{i}(z)$
describes the probability of fragmentation of the parton $i$ into
a heavy meson. For $D$- and $B$-mesons production, as well as for non-prompt $J/\psi$ production, 
the corresponding fragmentation functions are known from the literature~~\cite{Binnewies:1998vm,Kniehl:1999vf,Kneesch:2007ey}.
While in~(\ref{eq:fragConvolution}) there is a sum over all parton
flavors, the dominant contribution to all the mentioned states stems
from the heavy $c$- and $b$-quarks. This implies that the cross-section
$d\sigma_{pp\to\bar{Q}_{i}Q_{i}+X}/d\eta\,d^{2}p_{T}$, for heavy
quark production might be evaluated in the heavy quark mass limit.
It is convenient to separate explicitly the leptonic and hadronic
parts of the cross-section, and rewrite it as~\cite{Zyla:2020zbs,Rezaeian:2012ji}
\begin{equation}
\frac{d\sigma_{ep\to\bar{Q}_{i}Q_{i}+X}}{d\eta d^{2}p_{T}}=\frac{\alpha_{{\rm em}}Q^{2}}{\left(s_{ep}-m^{2}\right)\pi}\,\left[\left(1-y\right)\frac{d\sigma_{L}}{d\eta\,d^{2}p_{T}}+\left(1-y+\frac{y^{2}}{2}\right)\frac{d\sigma_{T}}{d\eta\,d^{2}p_{T}}\right],\label{eq:LTSep}
\end{equation}
where $d\sigma_{L}$ and $d\sigma_{T}$ in the right-hand side of
the equation~(\ref{eq:LTSep}) correspond to the cross-sections of
heavy quark production by a longitudinally and transversely polarized
photon respectively. In the literature the results for leptonic processes
are frequently discussed in terms of these photon-proton cross-sections
$d\sigma_{L,\,T}$, which have simpler structure. In the dipole approach
the cross-sections $d\sigma_{L,T}$ are given by
\begin{align}
\frac{d\sigma_{a}}{d\eta\,d^{2}p_{T}} & =\int_{0}^{1}dz\int\frac{d^{2}r_{1}}{4\pi}\,\int\frac{d^{2}r_{2}}{4\pi}e^{i\left(\boldsymbol{r}_{1}-\boldsymbol{r}_{2}\right)\cdot\boldsymbol{k}_{T}}\times\label{eq:QQ}\\
 & \,\times\Psi_{a}^{\dagger}\left(r_{2},\,z\right)\Psi_{a}^{\dagger}\left(r_{1},\,z\right)N_{M}\left(x_{2}(y);\,\vec{r}_{1},\,\vec{r}_{2}\right),\quad a=L,\,T\nonumber 
\end{align}
where $\eta$ and $\boldsymbol{p}_{T}$ are the rapidity and transverse
momenta of the produced heavy meson; $\Psi_{a}(r,\,z)$ is the $\bar{Q}Q$
component of the light-cone wave function of the photon; $\boldsymbol{r}_{1,2}$
are the transverse separation between quarks in the amplitude and
its conjugate; while $z$ is the light-cone fraction of the photon momentum
carried by the quark. For $\Psi_{a}$, in the heavy quark mass limit
we may use the standard perturbative expressions~\cite{Dosch:1996ss,Bjorken:1970ah}
\begin{align}
\Psi_{T}^{\dagger}\left(r_{2},\,z,\,Q^{2}\right)\Psi_{T}\left(r_{1},\,z,\,Q^{2}\right) & =\frac{\alpha_{s}N_{c}}{2\pi^{2}}\left\{ \epsilon_{f}^{2}\,K_{1}\left(\epsilon_{f}r_{1}\right)K_{1}\left(\epsilon_{f}r_{2}\right)\left[e^{i\theta_{12}}\,z^{2}+e^{-i\theta_{12}}(1-z)^{2}\right]\right.\\
 & \left.+m_{f}^{2}K_{0}\left(\epsilon_{f}r_{1}\right)K_{0}\left(\epsilon_{f}r_{2}\right)\right\} ,\nonumber \\
\Psi_{L}^{\dagger}\left(r_{2},\,z,\,Q^{2}\right)\Psi_{L}\left(r_{1},\,z,\,Q^{2}\right) & =\frac{\alpha_{s}N_{c}}{2\pi^{2}}\,\left\{ 4Q^{2}z^{2}(1-z)^{2}K_{0}\left(\epsilon_{f}r_{1}\right)K_{0}\left(\epsilon_{f}r_{2}\right)\right\} ,
\end{align}
where $\theta_{12}$ is the azimuthal angle between vectors $\boldsymbol{r}_{1}$
and $\boldsymbol{r}_{1}$, $m_{f}$ is the mass of the quark of flavor
$f$, and we used standard shorthand notations
\begin{equation}
\epsilon_{f}^{2}=z\,(1-z)\,Q^{2}+m_{f}^{2},
\end{equation}
\begin{equation}
\left|\Psi^{(f)}\left(r,\,z,\,Q^{2}\right)\right|^{2}=\left|\Psi_{T}^{(f)}\left(r,\,z,\,Q^{2}\right)\right|^{2}+\left|\Psi_{L}^{(f)}\left(r,\,z,\,Q^{2}\right)\right|^{2}.
\end{equation}

\begin{figure}
\includegraphics[width=6.5cm]{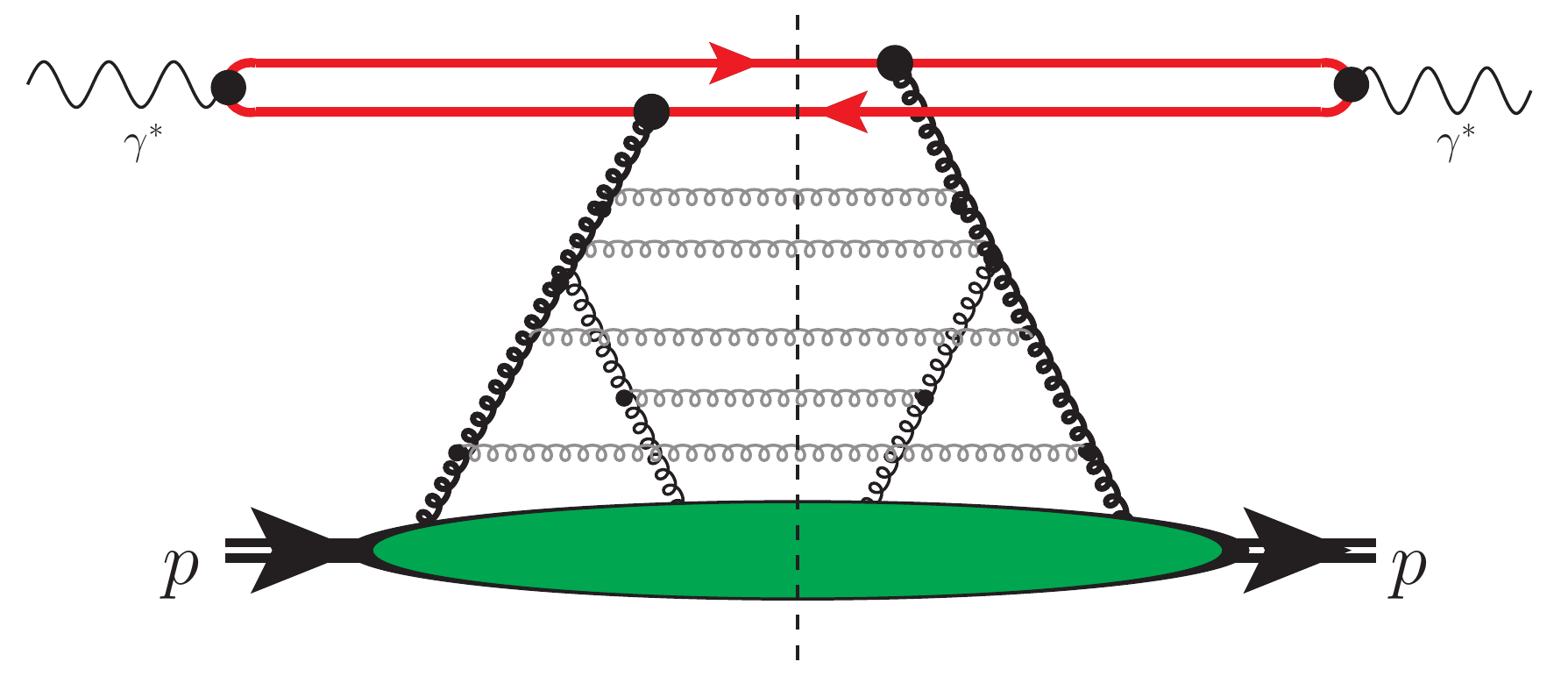}\includegraphics[width=12cm]{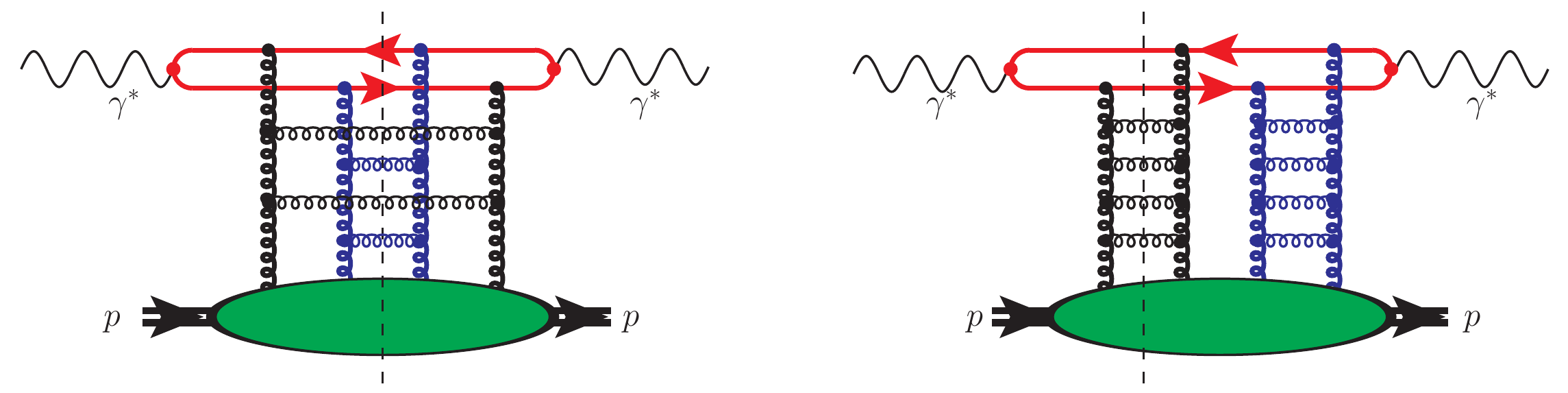}

\caption{\label{fig:DipoleCrossSections-2Pom}Left plot: the dominant contribution
to electroproduction of heavy quark pairs (single-pomeron contribution).
The dipole amplitude~\cite{Iancu:2003ge,RESH,Kowalski:2006hc,Watt:2007nr},
being the solution of the Balitsky-Kovchegov (BK) equation, effectively
includes all possible fan-like contributions shown by the horizontal grey
lines (resummation of all possible fan-like topologies is implied).
Central and right plots: possible higher-twist contributions due
to multipomeron (two-pomeron) mechanisms. For the sake of legibility
the fan-like structures were simplified down to simple gluon ladders
(as in BFKL). The difference in the number of cut pomerons in the central
and right plots will lead to a difference of multiplicity distributions.
In all plots the vertical dashed grey line stands for the unitarity
cut, the blob in the lower part is the hadronic target (proton); the
fermionic loop in the upper part of the figure includes a summation
over all possible gluons.}
\end{figure}

\begin{figure}
\includegraphics[width=9cm]{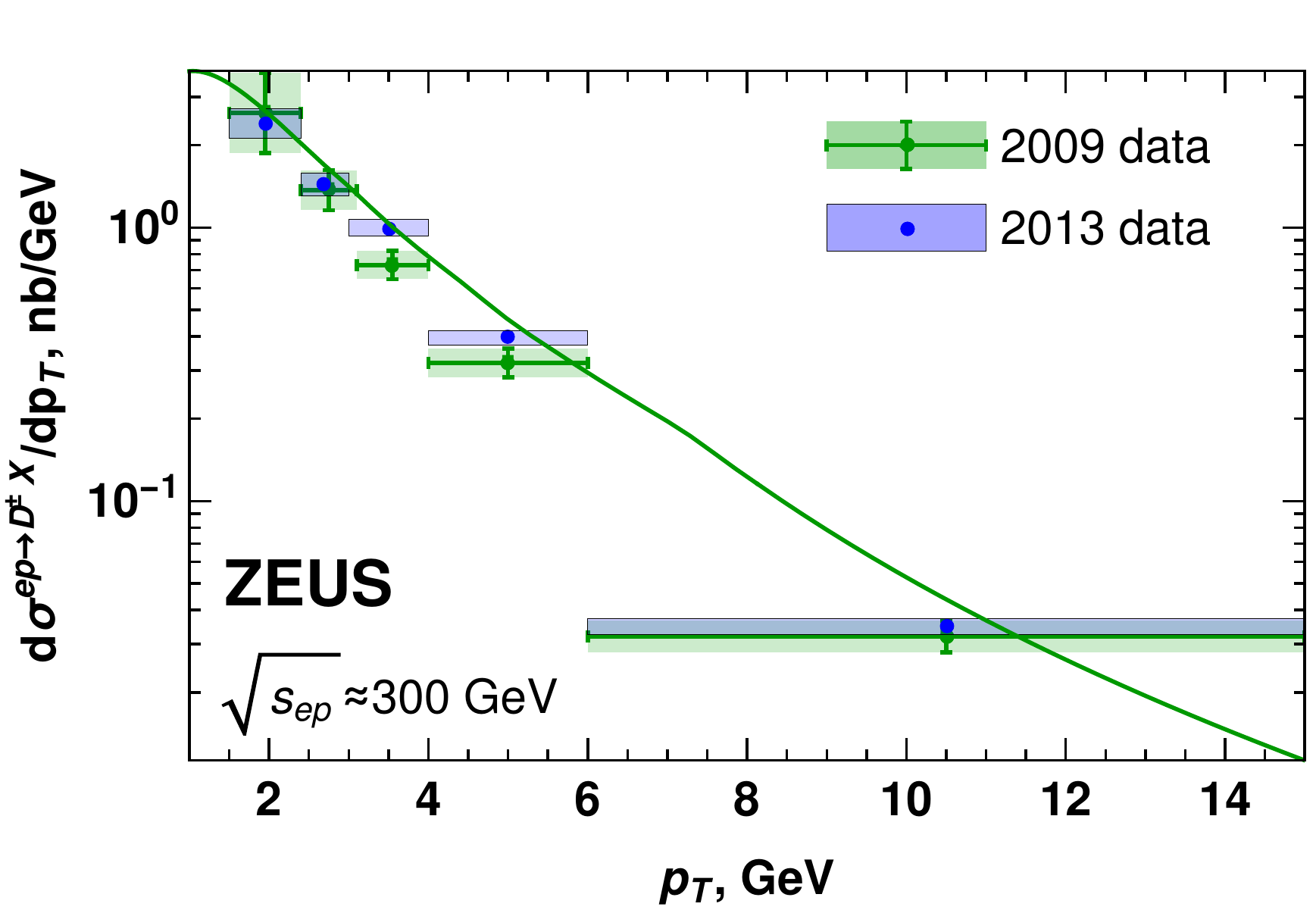}\includegraphics[width=9cm]{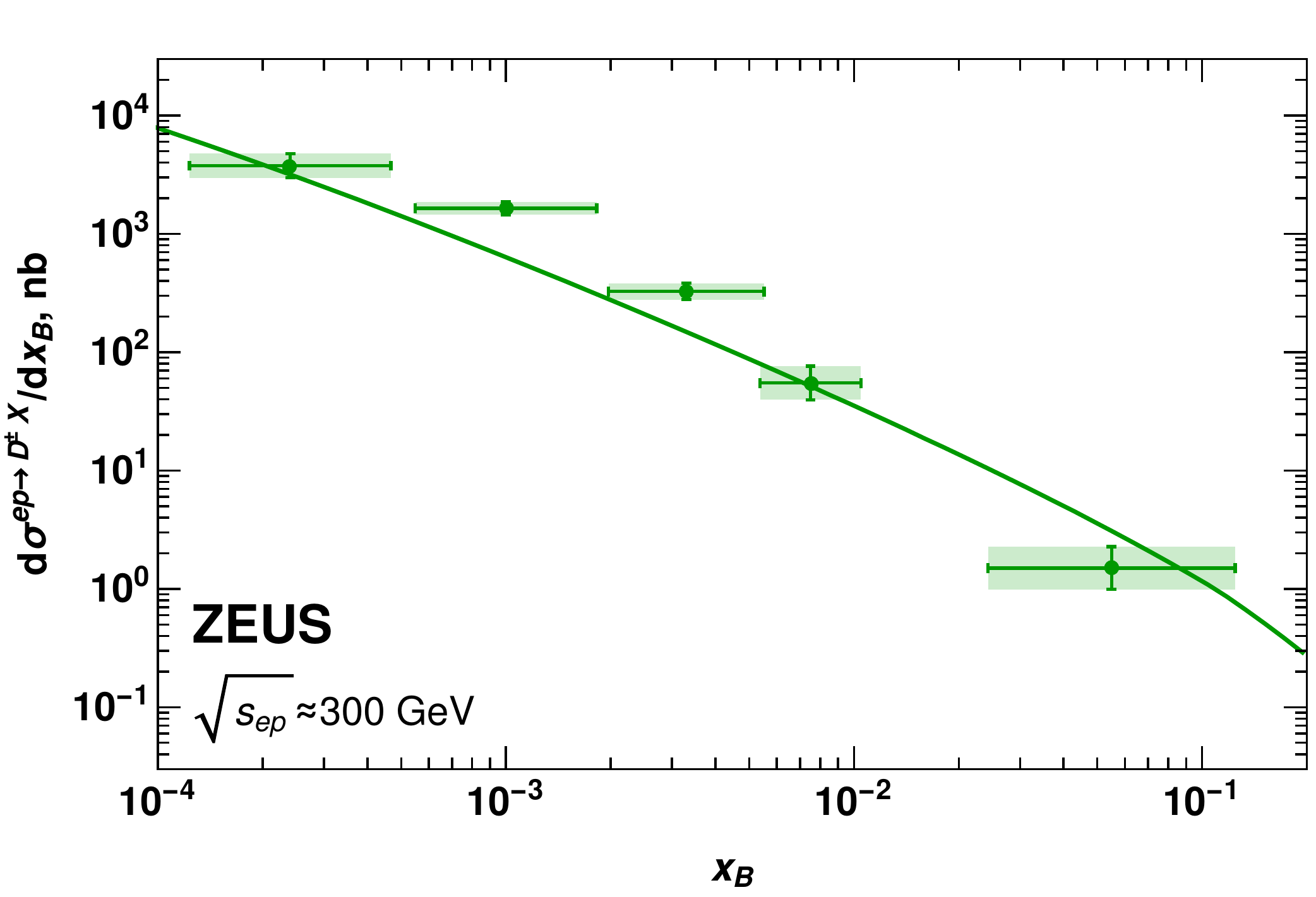}

\caption{\label{fig:pTDependence-1Pom}The $p_{T}$- and $x_{B}$-dependence
of the $D$-meson production cross-section for $D^{+}$-mesons in
the leading twist (single pomeron) contribution. The experimental
data are from~\cite{Acharya:2017jgo,Acharya:2019mgn,Aaij:2013mga}.
For $D^{0}$-mesons the dependence on the kinematical variables $p_{T},\,x_{B}$
has a similar shape, although it differs by a numerical factor of two.}
\label{Diags_DMesons-2} 
\end{figure}

\begin{figure}
\includegraphics[width=8.5cm]{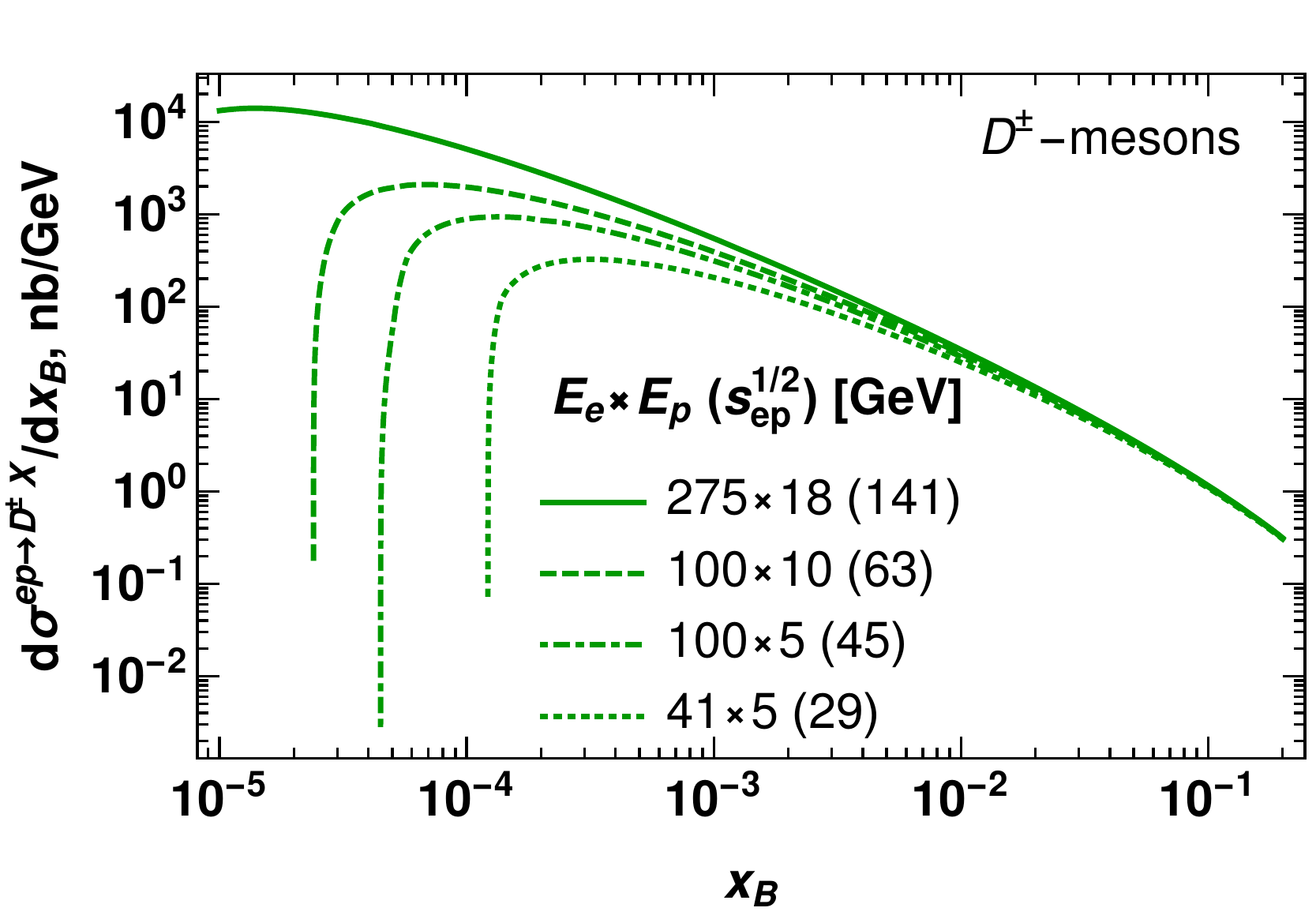}\includegraphics[width=8.5cm]{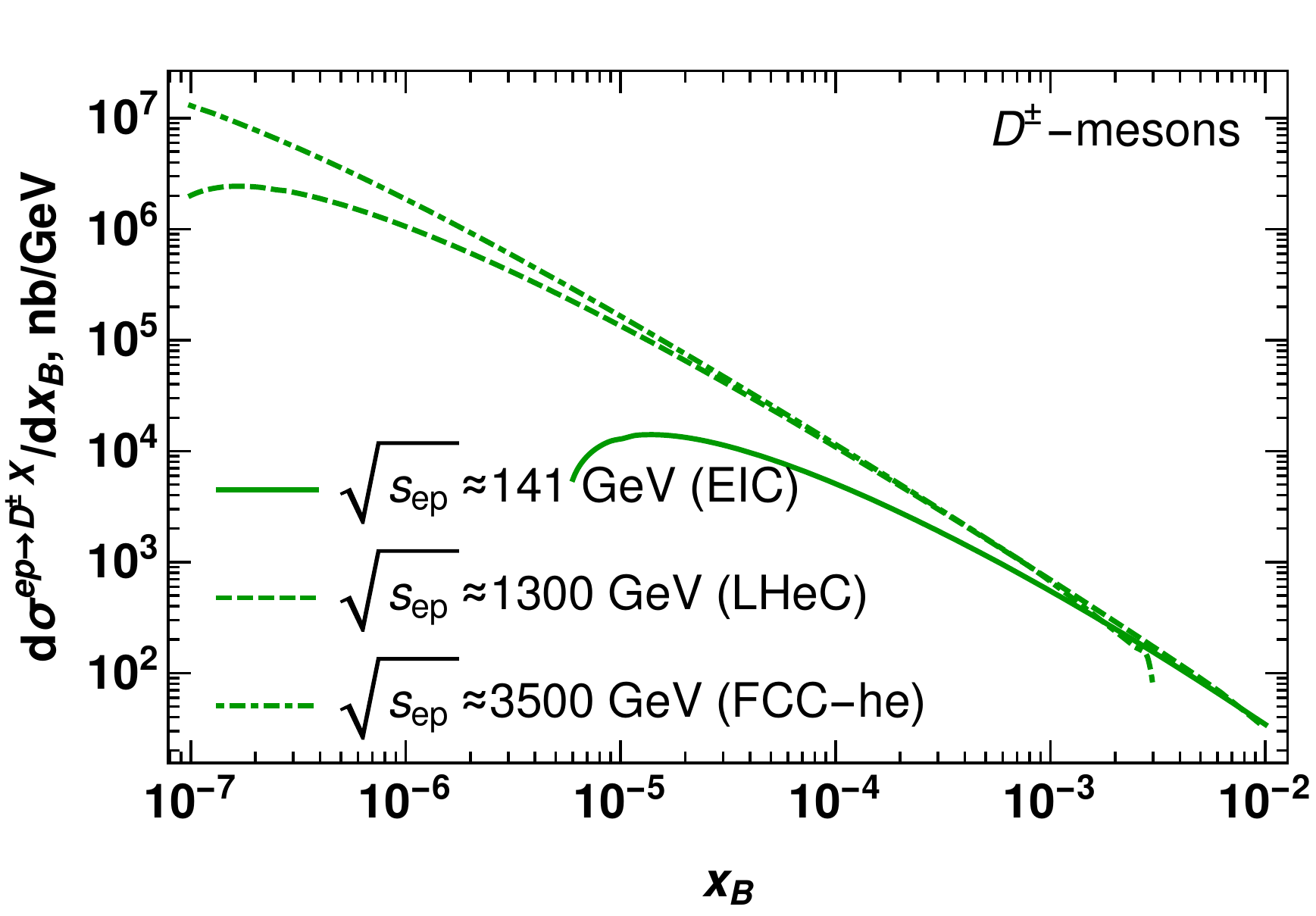}

\includegraphics[width=8.5cm]{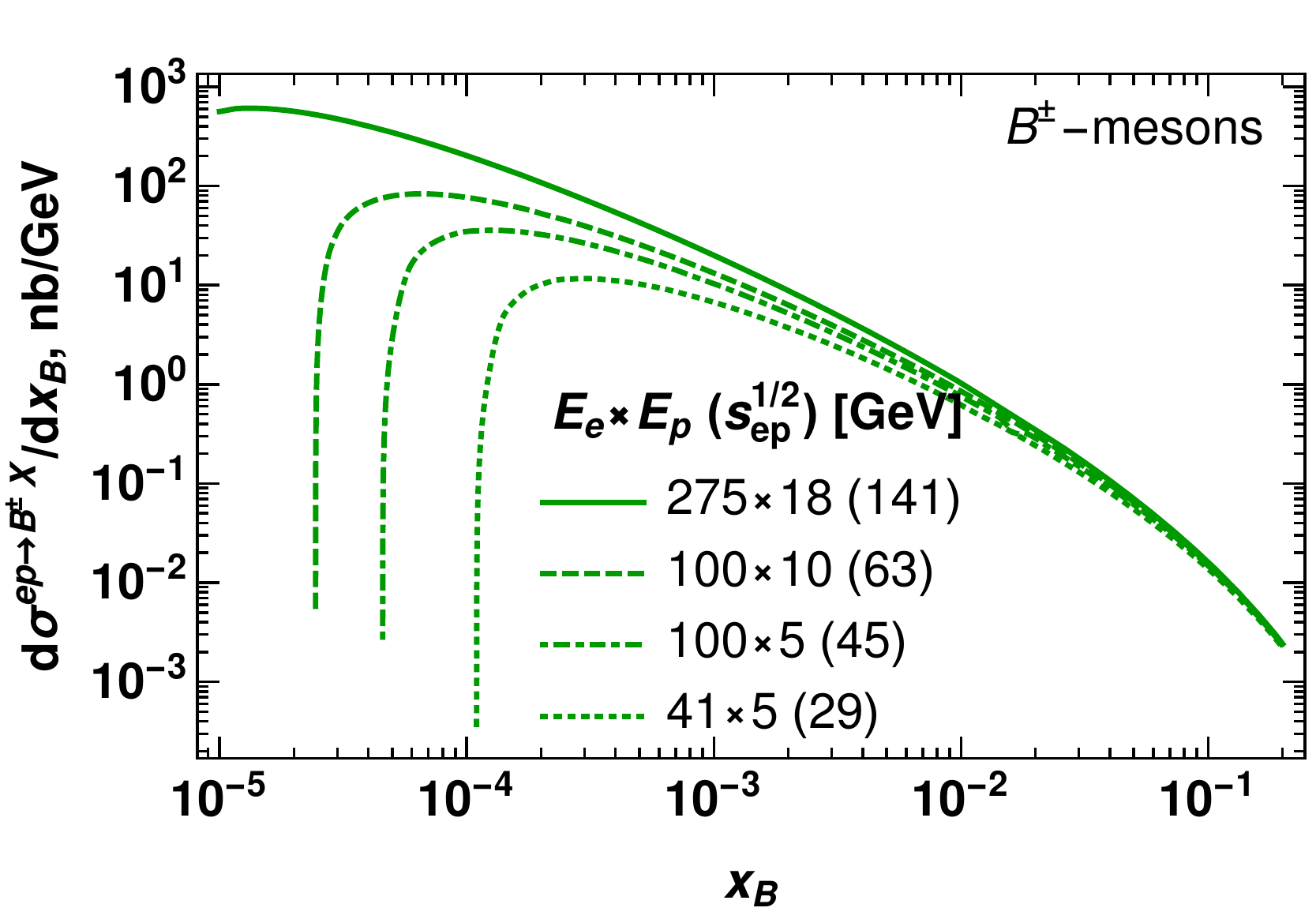}\includegraphics[width=8.5cm]{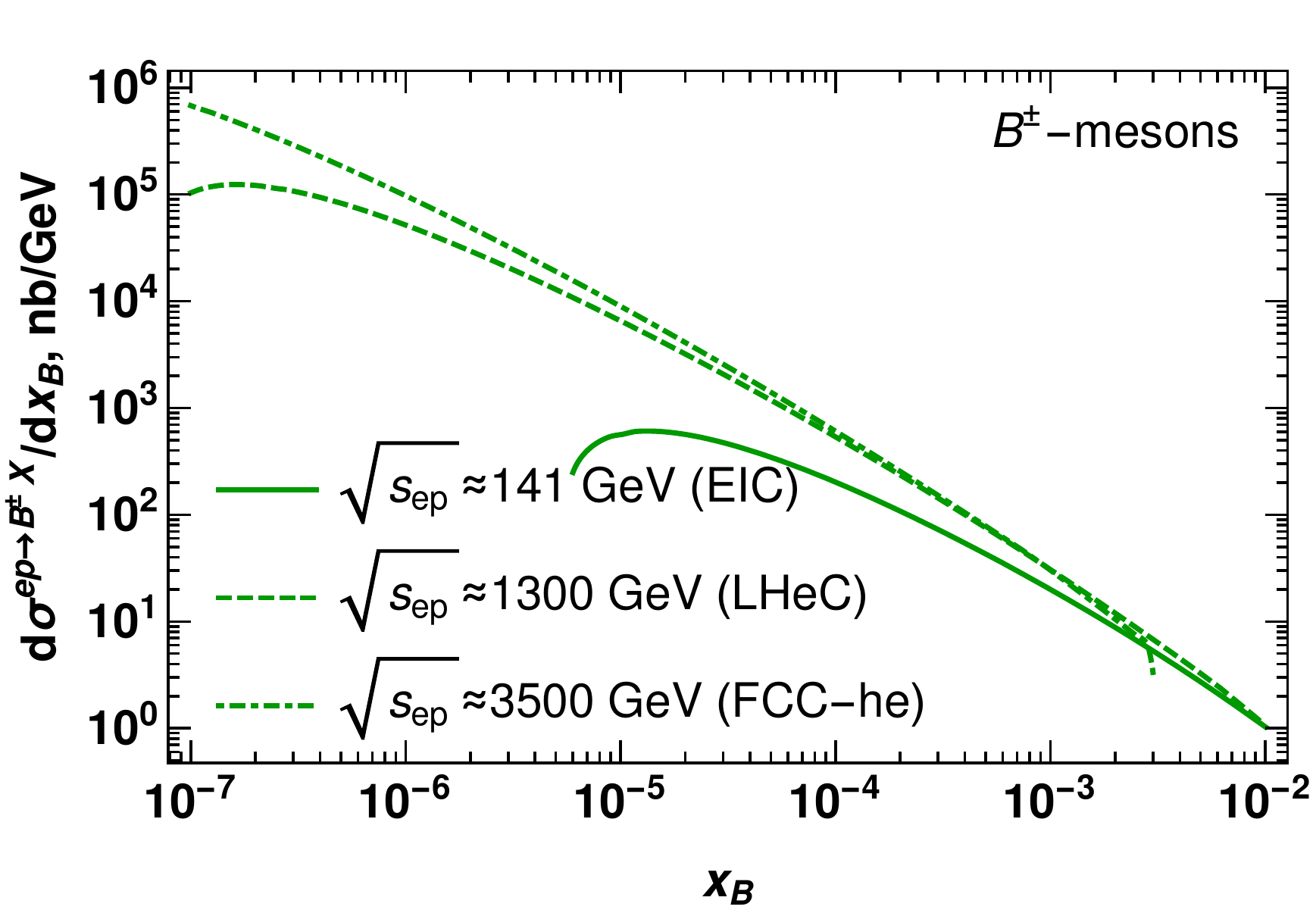}

\includegraphics[width=8.5cm]{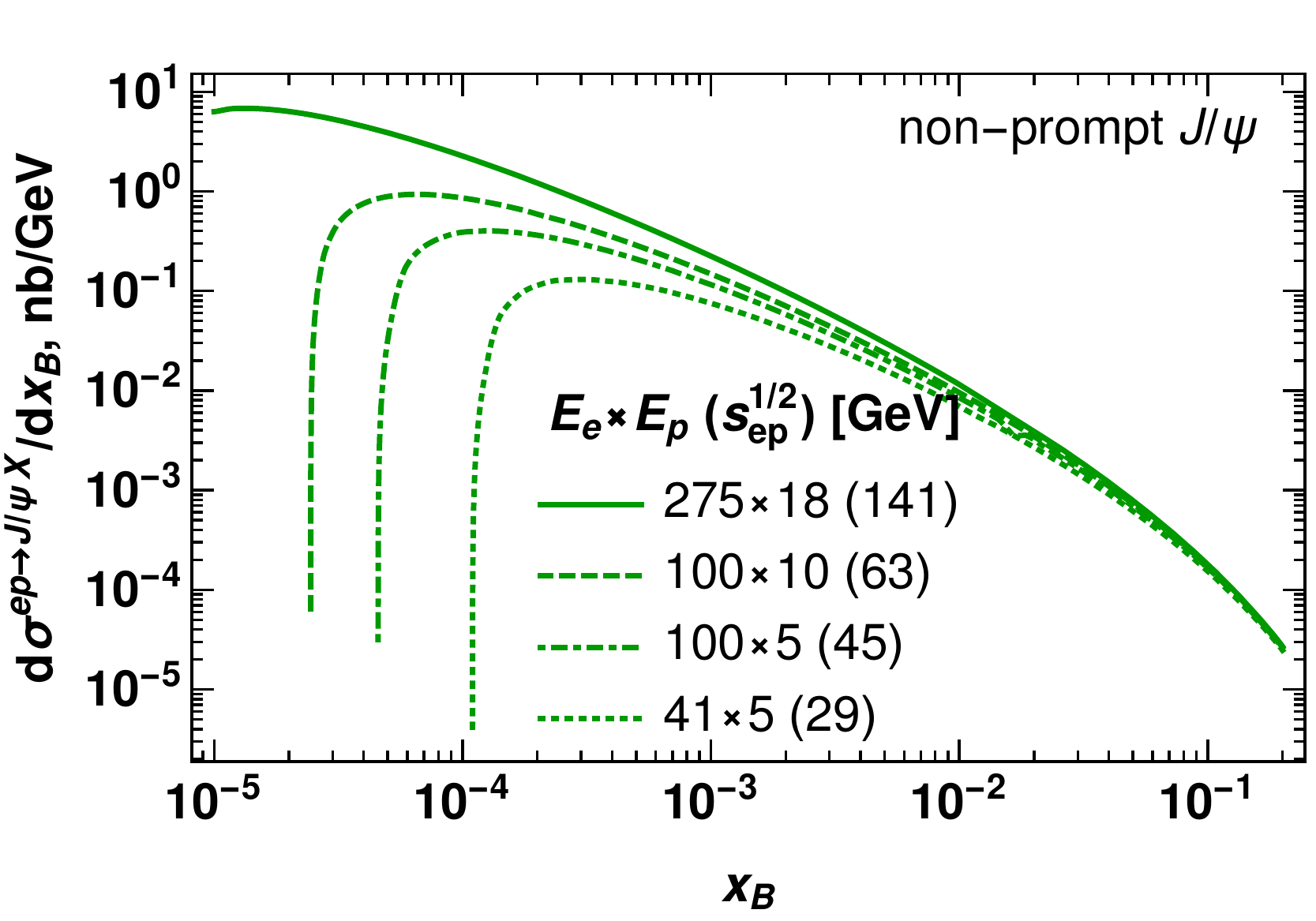}\includegraphics[width=8.5cm]{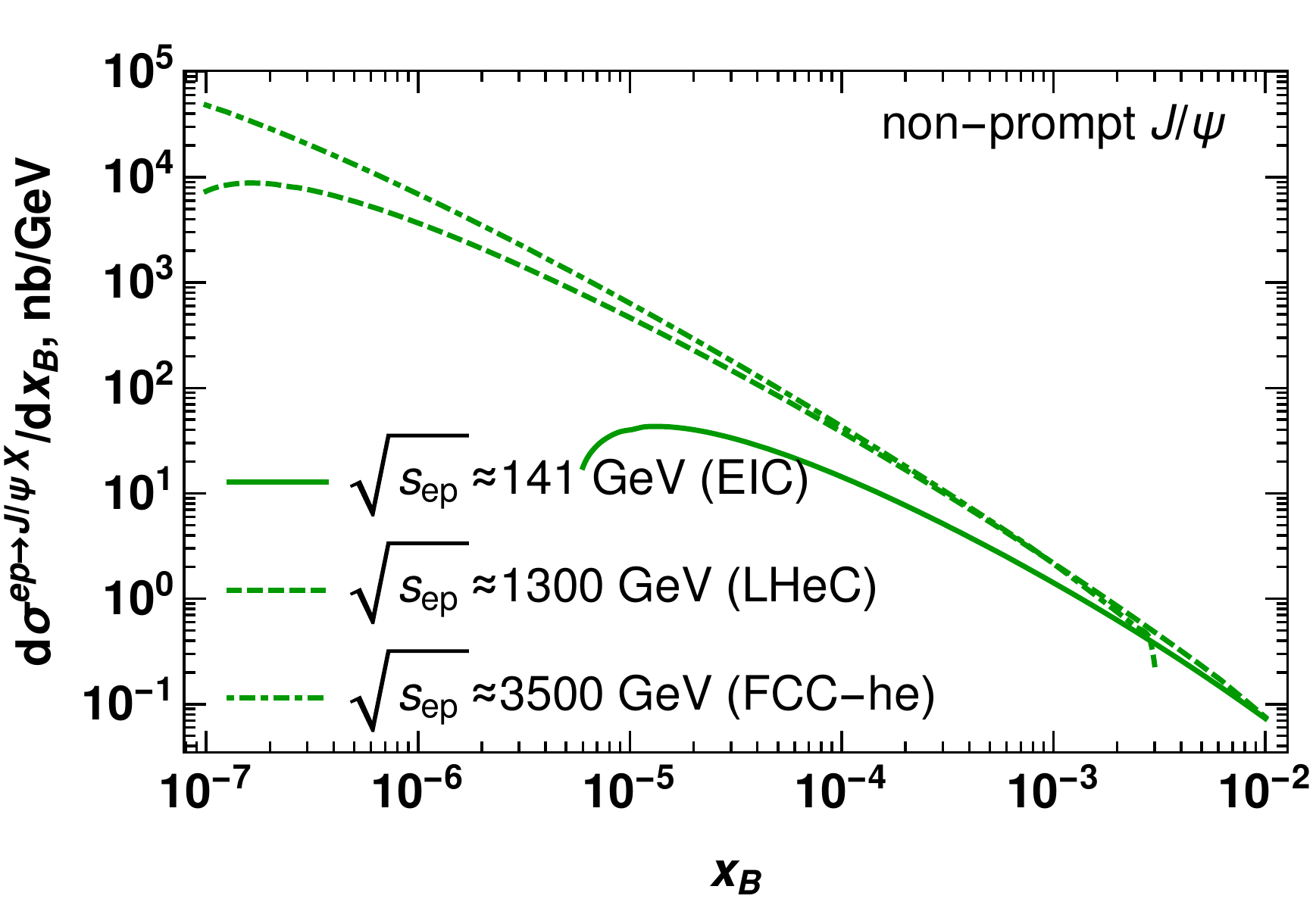}

\caption{\label{fig:xBjDependence-1Pom-Future}The $x_{B}$-dependence of the
production cross-section for $D$-mesons (upper row) $B$-mesons (central
row) and non-prompt $J/\psi$ mesons (lower row). The left column
corresponds to different energy sets in the kinematics of the future
EIC; the right column corresponds to predictions for LHeC and FCC-he
accelerators. For the sake of brevity we consider only charged $D^{+}$
and $B^{+}$-mesons; for other $D$- and $B$-mesons the $x_{B}$-dependence
has similar shape, although it differs by a numerical factor of two.}
\end{figure}

\begin{figure}
\includegraphics[width=9cm]{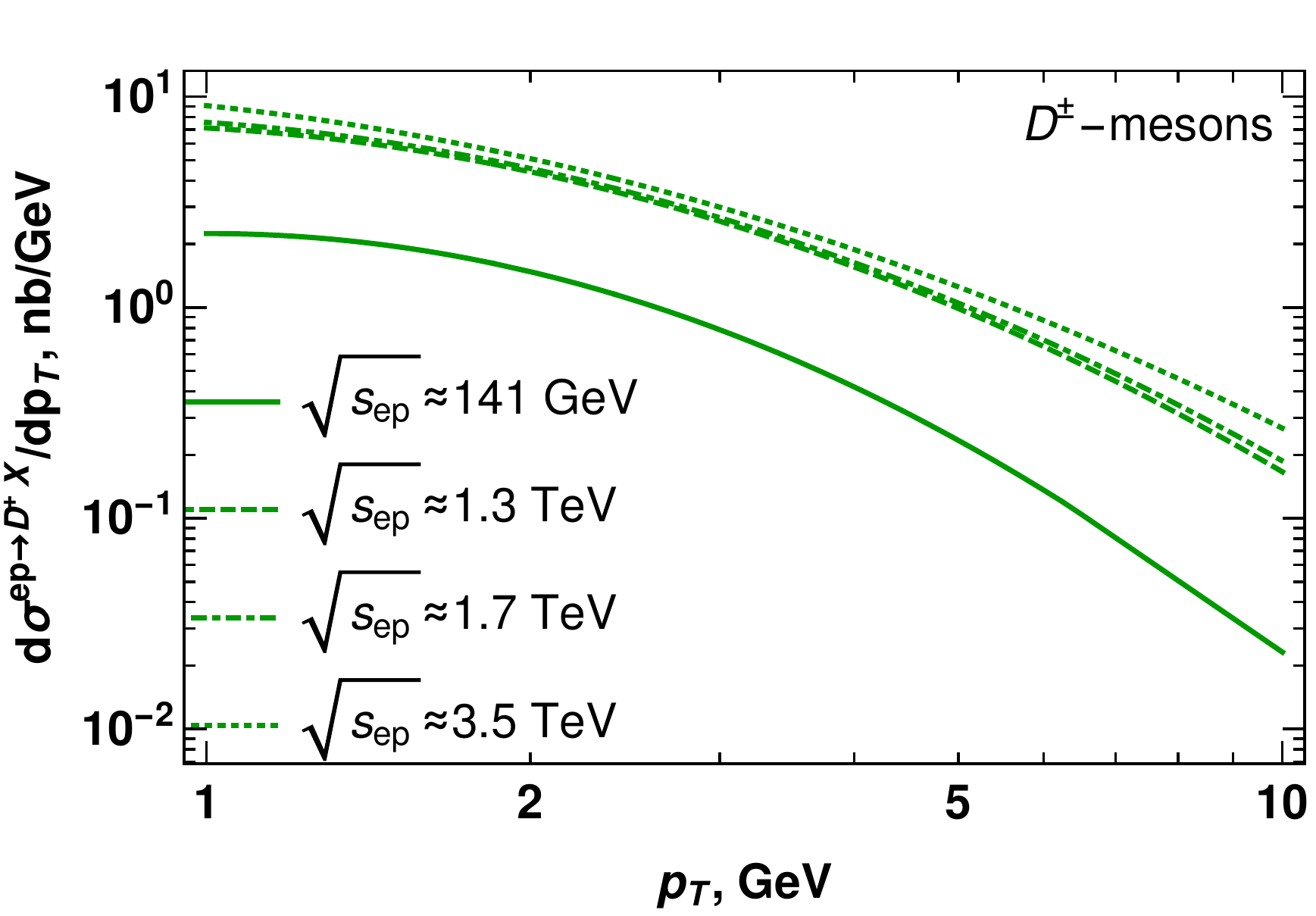}b\includegraphics[width=9cm]{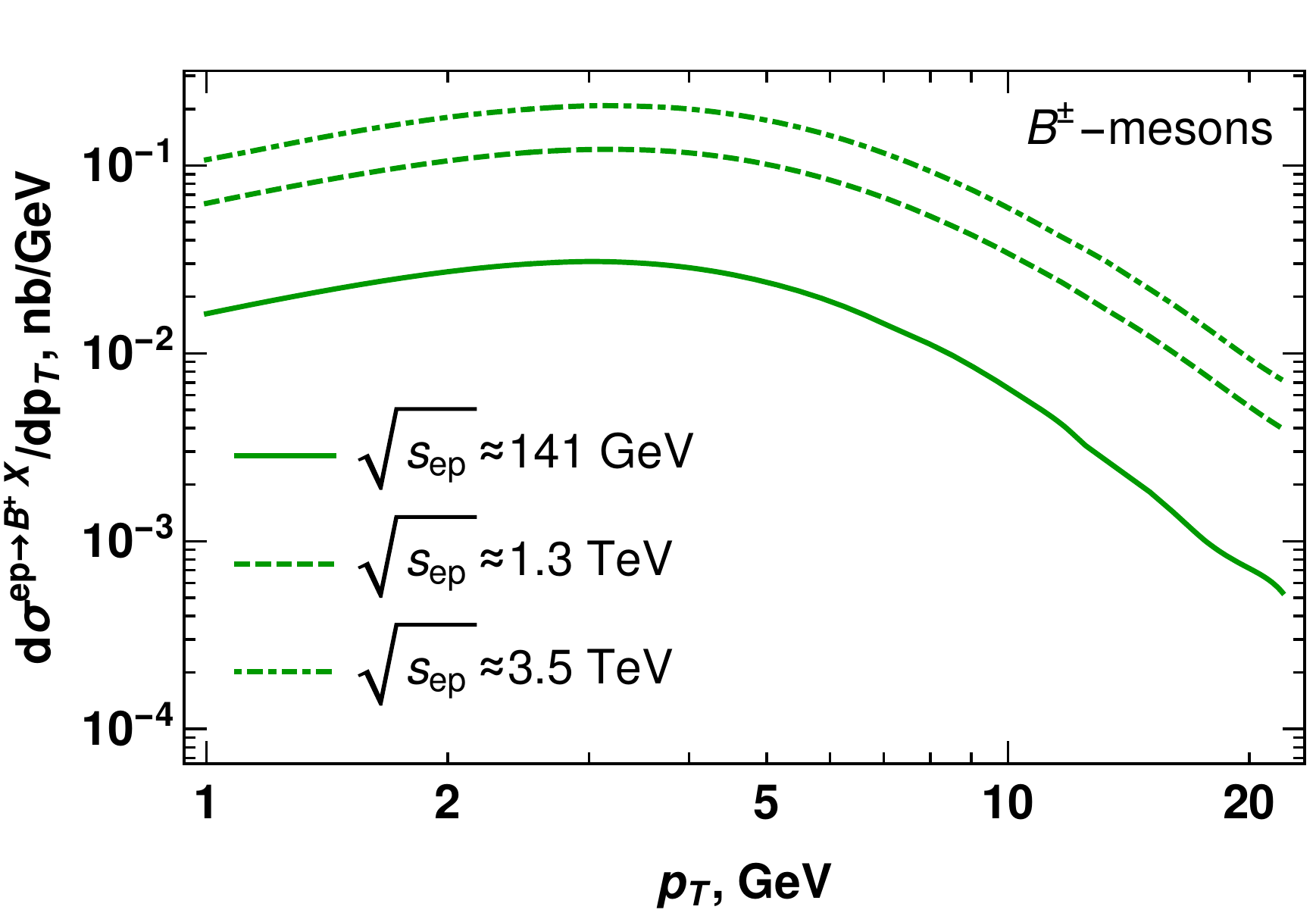}

\caption{\label{fig:pTDependence-1Pom-Future}The $p_{T}$-dependence of the
$D^{\pm}$- and $B^{\pm}$-meson production cross-section $d\sigma/dp_{T}$
in the kinematics of the future accelerators EIC ($\sqrt{s}_{ep}\approx$141
GeV), LHeC ($\sqrt{s}_{ep}\approx1.3-1.7$ TeV) and FCC-he~($\sqrt{s}_{ep}\approx3.5$
TeV). The difference between the shapes of the $D$- and $B$-mesons in
the small-$p_{T}$ kinematics stems from the difference of masses
of the $b$ and $c$ quarks. For the sake of definiteness we consider
only charged mesons; for other $D$- and $B$-mesons the $p_{T}$-dependence
has a similar shape, although it differs by a numerical factor of two.}
\end{figure}

The meson production amplitude $N_{M}$ depends on the mechanism of
the $Q\bar{Q}$ pair formation. For the case of production on a single-pomeron
(see the left panel of the Figure~\ref{fig:DipoleCrossSections-2Pom}),
in leading order it is given by by~\cite{Kopeliovich:2002yv,Goncalves:2017chx}
\begin{eqnarray}
 &  & N_{M}^{(1)}\left(x,\,\,\vec{\boldsymbol{r}}_{1},\,\vec{\boldsymbol{r}}_{2}\right)=\label{eq:N2-1}\\
 &  & =-\frac{1}{2}N\left(x,\,\vec{\boldsymbol{r}}_{1}-\vec{\boldsymbol{r}}_{2}\right)-\frac{1}{16}\left[N\left(x,\,\vec{\boldsymbol{r}}_{1}\right)+N\left(x,\,\vec{\boldsymbol{r}}_{2}\right)\right]-\frac{9}{8}N\left(x,\,\bar{z}\left(\vec{\boldsymbol{r}}_{1}-\vec{\boldsymbol{r}}_{2}\right)\right)\nonumber \\
 &  & +\frac{9}{16}\left[N\left(x,\,\bar{z}\vec{\boldsymbol{r}}_{1}-\vec{\boldsymbol{r}}_{2}\right)+N\left(x,\,\bar{z}\vec{\boldsymbol{r}}_{2}-\vec{\boldsymbol{r}}_{1}\right)+N\left(x,\,\bar{z}\vec{\boldsymbol{r}}_{1}\right)+N\left(x,\,\bar{z}\vec{\boldsymbol{r}}_{2}\right)\right],\nonumber 
\end{eqnarray}
where $N(x,\,r)$ is the amplitude of the color singlet dipole scattering.
The amplitude~(\ref{eq:N2-1}) has a structure similar to the leading
twist result for the \emph{hadro}production of heavy quarks; however,
this similarity is no longer valid for higher twist amplitudes. For
numerical estimates of this contribution, we need to fix a parametrization
of the amplitude $N(x,\,r)$. In what follows, for the sake of definiteness
we will use the CGC parametrization of the dipole amplitude, which was
proposed in~\cite{Iancu:2003ge} (see also~\cite{Kowalski:2003hm,Kowalski:2006hc,Watt:2007nr,RESH}
for more recent phenomenological analyses). Since we are interested
in the $p_{T}$ dependence, we will use the impact parameter dependent fit, taken
from~\cite{RESH}. As we can see from Figure~\ref{fig:pTDependence-1Pom},
the single-pomeron contribution provides a very reasonable description
of the available data from HERA. In Figures~\ref{fig:xBjDependence-1Pom-Future},~\ref{fig:pTDependence-1Pom-Future}
we have shown the theoretical expectations for the cross-sections
of $D^{\pm}$-, $B^{\pm}$- and non-prompt $J/\psi$ meson production,
in the kinematics of the future accelerators EIC ($\sqrt{s}_{ep}$
up to 141 GeV), LHeC ($\sqrt{s}_{ep}\approx1.3$ TeV) and FCC-he~($\sqrt{s}_{ep}\approx3.5$
TeV)~\cite{Accardi:2012qut,AbdulKhalek:2021gbh,AbelleiraFernandez:2012cc,Mangano:2017tke,Agostini:2020fmq,Abada:2019lih}.

It is also interesting to understand the role of the multipomeron
mechanisms in electroproduction. While sometimes it is assumed that
all such contributions are taken into account by the universal dipole
cross-section, in reality the situation is more complicated. The CGC
parametrization~\cite{Kowalski:2006hc,Watt:2007nr,RESH}, used for
our numerical estimates, does not take into account such corrections. 
Another widely used parametrization of the dipole cross-section,
the so-called $b$-Sat~\cite{Kowalski:2003hm,Rezaeian:2012ji}, takes
into account such corrections, making additional simplifying assumptions.
For this reason our goal is to perform a microscopic evaluation using
the CGC model. We understand that a systematic evaluation of all such
corrections in high-multiplicity events presents a challenging problem,
and for this reason we will focus on the contribution of two-pomeron mechanisms,
which are shown in the central and right panels of the Figure~\ref{fig:DipoleCrossSections-2Pom}.
Formally such contributions are expected to be small, because they
are of higher twist. However, it is desired to reassess them for electroproduction,
because earlier studies~\cite{Schmidt:2020fgn} revealed that for
\emph{hadro}production such corrections might be pronounced in the charm
sector and in small-$p_{T}$ kinematics, especially for high-multiplicity
events. In what follows we will refer to the diagrams shown in the
central and right panels of the Figure~\ref{fig:DipoleCrossSections-2Pom}
as genuine and interference corrections (in view of the clear interference
nature of the latter). For both types of contributions the corresponding
cross-section has the familiar structure~(\ref{eq:QQ}), so these
corrections might be rewritten as an additional contribution to the
amplitude $N_{M}$ given by
\begin{eqnarray}
 &  & N_{M}^{(2)}\left(x,\,\,\vec{r}_{1},\,\vec{r}_{2}\right)=N_{M}^{({\rm genuine})}\left(x,\,z,\,\vec{\boldsymbol{r}}_{1},\,\vec{\boldsymbol{r}}_{2}\right)+N_{M}^{({\rm int})}\left(x,\,z,\,\vec{\boldsymbol{r}}_{1},\,\vec{\boldsymbol{r}}_{2}\right)\label{eq:N3}
\end{eqnarray}
where
\begin{align}
N_{M}^{({\rm genuine})}\left(x,\,z,\,\vec{\boldsymbol{r}}_{1},\,\vec{\boldsymbol{r}}_{2}\right)\approx & \,\frac{1}{8}\left[N_{+}^{2}\left(x,\,z,\,\vec{\boldsymbol{r}}_{1},\,\vec{\boldsymbol{r}}_{2}\right)\left(\frac{3N_{c}^{2}}{8}\right)+N_{-}^{2}\left(x,\,\vec{\boldsymbol{r}}_{1},\,\vec{\boldsymbol{r}}_{2}\right)\left(\frac{\left(43\,N_{c}^{4}-320N_{c}^{2}+720\right)}{72\,N_{c}^{2}}\right)\right.\label{eq:N3Direct}\\
 & \qquad+\left.\frac{\left(N_{c}^{2}-4\right)}{2}N_{+}\left(x,\,z,\,\vec{\boldsymbol{r}}_{1},\,\vec{\boldsymbol{r}}_{2}\right)N_{-}\left(x,\,\vec{\boldsymbol{r}}_{1},\,\vec{\boldsymbol{r}}_{2}\right)\right],\nonumber 
\end{align}
\begin{align}
N_{M}^{({\rm int})}\left(x,\,z,\,\vec{\boldsymbol{r}}_{1},\,\vec{\boldsymbol{r}}_{2}\right)= & -\,\frac{3}{16}\left[2\,N_{+}\left(x,\,z,\,\vec{\boldsymbol{r}}_{1},\,\vec{\boldsymbol{r}}_{2}\right)\tilde{N}_{+}\left(x,\,z,\,\vec{\boldsymbol{r}}_{2}\right)\left(\frac{3N_{c}^{2}}{8}\right)+\right.\label{eq:N3Interf}\\
 & -N_{-}\left(z,\,\vec{\boldsymbol{r}}_{1},\,\vec{\boldsymbol{r}}_{2}\right)\tilde{N}_{-}\left(x,\,\vec{\boldsymbol{r}}_{2}\right)\left(\frac{\left(43\,N_{c}^{4}-320N_{c}^{2}+720\right)}{72\,N_{c}^{2}}\right)+\nonumber \\
 & +\left.\frac{\left(N_{c}^{2}-4\right)}{2}\left(N_{+}\left(z,\,\vec{\boldsymbol{r}}_{1},\,\vec{\boldsymbol{r}}_{2}\right)\tilde{N}_{-}\left(x,\,\vec{\boldsymbol{r}}_{2}\right)+\tilde{N}_{+}\left(x,\,\vec{\boldsymbol{r}}_{2}\right)N_{-}\left(z,\,\vec{\boldsymbol{r}}_{1},\,\vec{\boldsymbol{r}}_{2}\right)\right)\right]\nonumber 
\end{align}
and we introduced the shorthand notations
\begin{align}
N_{-}\left(x,\,\vec{\boldsymbol{r}}_{1},\,\vec{\boldsymbol{r}}_{2}\right) & \equiv-\frac{1}{2}\left[N\left(x,\,\vec{\boldsymbol{r}}_{2}-\vec{\boldsymbol{r}}_{1}\right)-N\left(x,\,\vec{\boldsymbol{r}}_{1}\right)-N\left(x,\,\vec{\boldsymbol{r}}_{2}\right)\right]\\
N_{+}\left(x,\,z,\,\vec{\boldsymbol{r}}_{1},\,\vec{\boldsymbol{r}}_{2}\right) & \equiv-\frac{1}{2}\left[N\left(x,\,\vec{\boldsymbol{r}}_{2}-\vec{\boldsymbol{r}}_{1}\right)+N\left(x,\,\vec{\boldsymbol{r}}_{1}\right)+N\left(x,\,\vec{\boldsymbol{r}}_{2}\right)\right]+N\left(x,\,\bar{z}\vec{\boldsymbol{r}}_{1}-\vec{\boldsymbol{r}}_{2}\right)+N\left(x,\,\bar{z}\vec{\boldsymbol{r}}_{1}\right)\\
 & +N\left(x,\,-\bar{z}\vec{\boldsymbol{r}}_{2}+\vec{\boldsymbol{r}}_{1}\right)+N\left(x,\,-\bar{z}\vec{\boldsymbol{r}}_{2}\right)-2N\left(x,\,\bar{z}\left(\vec{\boldsymbol{r}}_{1}-\vec{\boldsymbol{r}}_{2}\right)\right)\nonumber 
\end{align}
The derivation of these expressions is straightforward and follows
the procedures described in~\cite{Kopeliovich:2001ee,Kopeliovich:2002yv,Goncalves:2017chx,Schmidt:2020fgn}.
Both functions $N_{\pm}\left(z,\,\vec{\boldsymbol{r}}_{1},\,\vec{\boldsymbol{r}}_{2}\right)$
are invariant with respect to the permutation $\boldsymbol{r}_{1}\leftrightarrow\boldsymbol{r}_{2}$.
The $p_{T}$-integrated cross-sections get contributions only from
$\vec{\boldsymbol{r}}_{1}=\vec{\boldsymbol{r}}_{2}=\vec{\boldsymbol{r}}$,
so the cross-sections $N_{\pm}$ simplify to 
\begin{align}
\tilde{N}_{-}\left(x,\,\,\vec{\boldsymbol{r}}\right) & \equiv N_{-}\left(x,\,\vec{\boldsymbol{r}},\,\vec{\boldsymbol{r}}\right)=N\left(x,\,\vec{\boldsymbol{r}}\right)\\
\tilde{N}_{+}\left(x,\,z,\,\vec{\boldsymbol{r}}\right) & \equiv N_{+}\left(x,\,z,\,\vec{\boldsymbol{r}},\,\vec{\boldsymbol{r}}\right)=2N\left(x,\,\bar{z}\vec{\boldsymbol{r}}\right)+2N\left(x,\,z\vec{\boldsymbol{r}}\right)-N\left(x,\,\vec{\boldsymbol{r}}\right)
\end{align}
In Figure~(\ref{fig:pTDependence-1Pom-Future}) we show
the ratio of cross-sections, where the numerator and denominator were
evaluated using the two-pomeron contribution~(\ref{eq:N3}) and the single-pomeron
contribution~(\ref{eq:N2-1}) respectively,
\begin{equation}
R\left(x_{B}\right)=\frac{d\sigma_{ep\to DX}^{(2)}/dx_{B}}{d\sigma_{ep\to DX}^{(1)}/dx_{B}}.\label{eq:R3}
\end{equation}
As we can see from Figure~(\ref{fig:pTDependence-1Pom-Future}),
in the kinematics of EIC the ratio is quite small. However, the situation
is different in the kinematics of the future LHeC and FCC-he colliders,
which might probe significantly smaller values of $x_{B}$. In that
kinematics the values of the two-pomeron contributions might reach
up to 40 per cent of the total result in the charm sector.

\begin{figure}
\includegraphics[width=9cm]{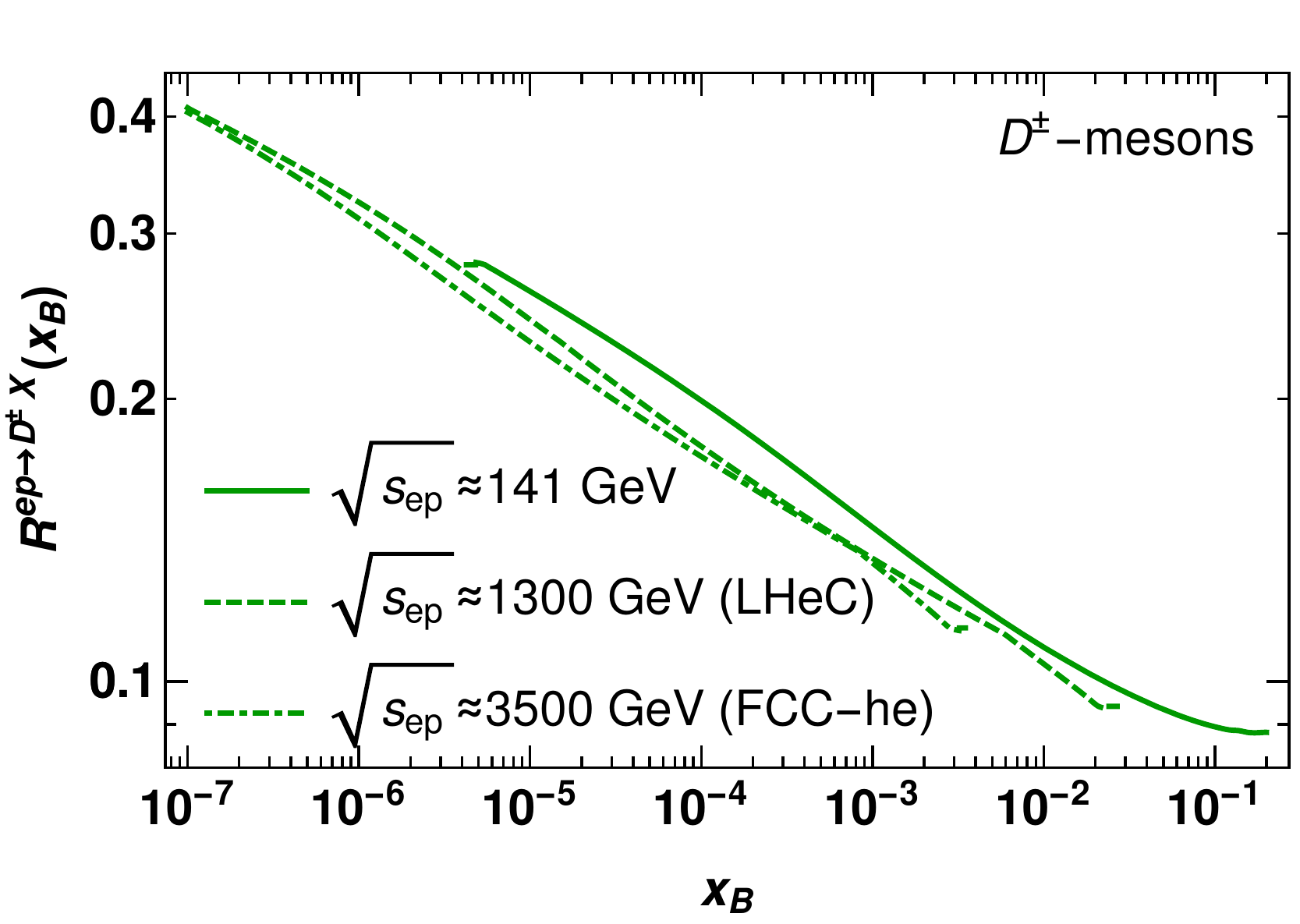}\includegraphics[width=9cm]{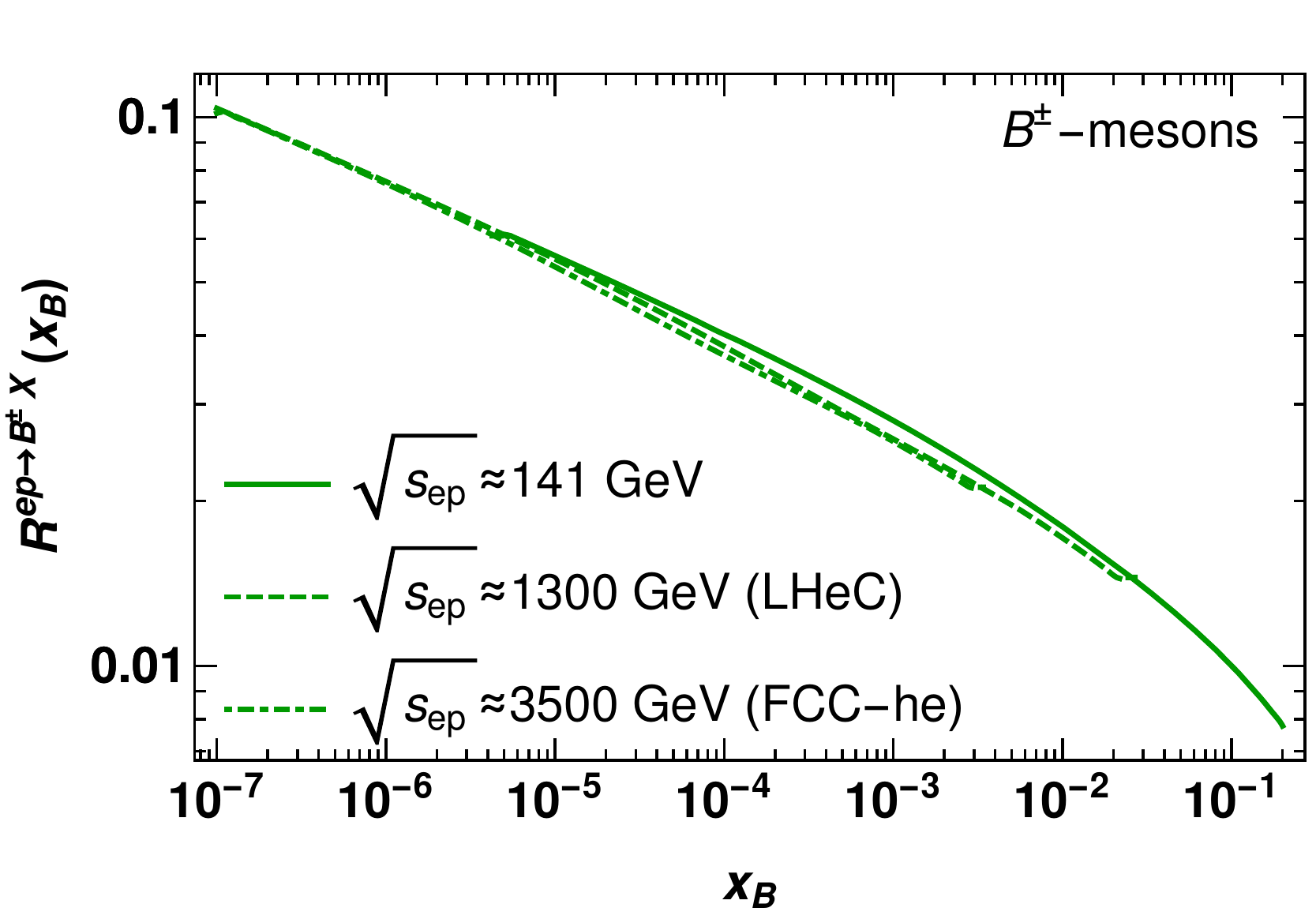}

\caption{\label{fig:Ratio-1Pom-Future-1}The ratio of the two-pomeron and single-pomeron
contributions, defined in~(\ref{eq:R3},) in the kinematics of future
accelerators. For the sake of definiteness we consider $D^{+}$ and
$B^{+}$-mesons; for other $D$- or $B$-mesons as well as for non-prompt
$J/\psi$ the ratio has a very similar shape. The region $x\lesssim5\times10^{-6}$
is kinematically forbidden for EIC energy, and for this reason the solid
curve abruptly vanishes there.}
\end{figure}

\section{Multiplicity dependence}

\label{sec:Numer}  The theoretical study of multiplicity dependence
in high energy collisions was initiated long ago in~\cite{Abramovsky:1973fm,Capella:1976ef,Bertocchi:1976bq,Shabelski:1977iv,Nikolaev:2006mx,Kaidalov:1982xe}
in the framework of the Regge approach. Relying on very general properties
of particle-reggeon vertices, which are largely independent of the underlying
quantum field theory, it was suggested that the enhanced multiplicity
of high energy final states could be considered as one of the manifestations
of the multiple pomeron contributions. Later it was demonstrated in~\cite{Bartels:1996hw,Bartels:2005wa,Kovchegov:1999yj,Kovchegov:2000hz,Kovchegov:2012mbw,Sjostrand:2004pf}
that all these findings are also valid in the context of QCD, and
thus could be confirmed by experimental evidence. The dependence on
multiplicity differs from the dependencies on other kinematic variables,
which are sometimes used for the extraction of dipole amplitudes, fragmentation
functions or parton distributions from experimental data. This implies
that the multiplicity dependence might be used as a litmus test to
probe the role of multipomeron contributions. 

The probability distribution $P\left(N_{{\rm ch}},\,\left\langle N_{{\rm ch}}\right\rangle \right)$
of high-multiplicity fluctuations inside each pomeron decreases rapidly
as a function of number of produced charged particles $N_{{\rm ch}}$,
as was found both at $ep$ and $pp$ collisions~\cite{Abelev:2012rz,Chekanov:2008ae}.
The theoretical modeling of the essentially nonperturbative probability
distribution $P\left(N_{{\rm ch}},\,\left\langle N_{{\rm ch}}\right\rangle \right)$
is very challenging. In order to exclude this common suppression factor,
it is convenient to analyze the multiplicity dependence of the \emph{ratio}
of two different processes. In $pp$ collisions usually the results
are presented for the ratio of cross-sections of heavy meson and inclusive
processes in a given multiplicity class, self-normalized to one for
$n\equiv N_{{\rm ch}}/\left\langle N_{{\rm ch}}\right\rangle =1$
for the sake of convenience \cite{Adam:2015ota,Thakur:2018dmp,Alice:2012Mult,Trzeciak:2015fgz,Ma:2016djk},
so effectively such ratios are proportional to a conditional probability
to measure a hadron provided $N_{{\rm ch}}$ charged particles are
produced in the final state. It was found that in charm sector such ratios
grow with multiplicity, which clearly indicates pronounced multipomeron
contributions. For $ep$ collisions this variable is not very convenient,
because multiparton configurations might contribute in a similar way
both to heavy meson productions and to the inclusive channel. In the latter
case a sizable contribution might come from large dipoles, for which
multipomeron contributions are even more pronounced than for heavy
mesons. Potentially the contribution of large dipoles might be suppressed
in some special kinematics (\emph{e.g}. at large virtualities $Q^{2}$);
however, it will be very challenging to measure multiplicity dependence
due to significantly smaller statistics. For this reason below we
will consider other variables which might present interest for studies
of multiplicity dependence. We need to mention that in contrast to
hadroproduction, the multiplicity dependence of electroproduction
is simpler at the conceptual level, because there are fewer different
mechanisms to produce an enhanced number of charged particles in the final
state.

The description of high-multiplicity events in the CGC/Sat framework
has been discussed in detail in~\cite{KOLEB,KLN,DKLN,Kharzeev:2000ph,Kovchegov:2000hz,LERE,Lappi:2011gu,Ma:2018bax}.
It is expected that at high multiplicities the dipole amplitude should
satisfy the same Balitsky-Kovchegov equation (and thus maintain its
form), although the saturation scale $Q_{s}(x,b)$, which contributes
to the dipole amplitude, should be modified as 
\begin{equation}
Q_{s}^{2}\left(x,\,b;\,n\right)\,\,=\,\,n\,Q^{2}\left(x,\,b\right).\label{QSN}
\end{equation}
For multipomeron configurations, we should take into account that
multiplicity fluctuations occur independently in each pomeron, and
the observed multiplicity $n$ might be shared in all possible ways
between all cut pomerons in a given rapidity window. However, as was
discussed in detail in~\cite{Levin:1993te,Levin:2018qxa,Siddikov:2019xvf},
with good precision we may assume that the observed multiplicity
$n$ is shared equally between all pomerons which participate in $ep$
process. Using this assumption, as well as certain convolution properties
of $P\left(N_{{\rm ch}},\,\left\langle N_{{\rm ch}}\right\rangle \right)$,
it is possible to show that for the ratio of different cross-sections
the probability distribution $P\left(N_{{\rm ch}},\,\left\langle N_{{\rm ch}}\right\rangle \right)$
cancels altogether. Thus for the evaluation of the cross-sections in a
given multiplicity class, we may use a simple prescription~(\ref{QSN}),
properly adjusting the parameter $n$ in each pomeron to take into account
equal sharing of total multiplicity.

\begin{figure}
\includegraphics[width=9cm]{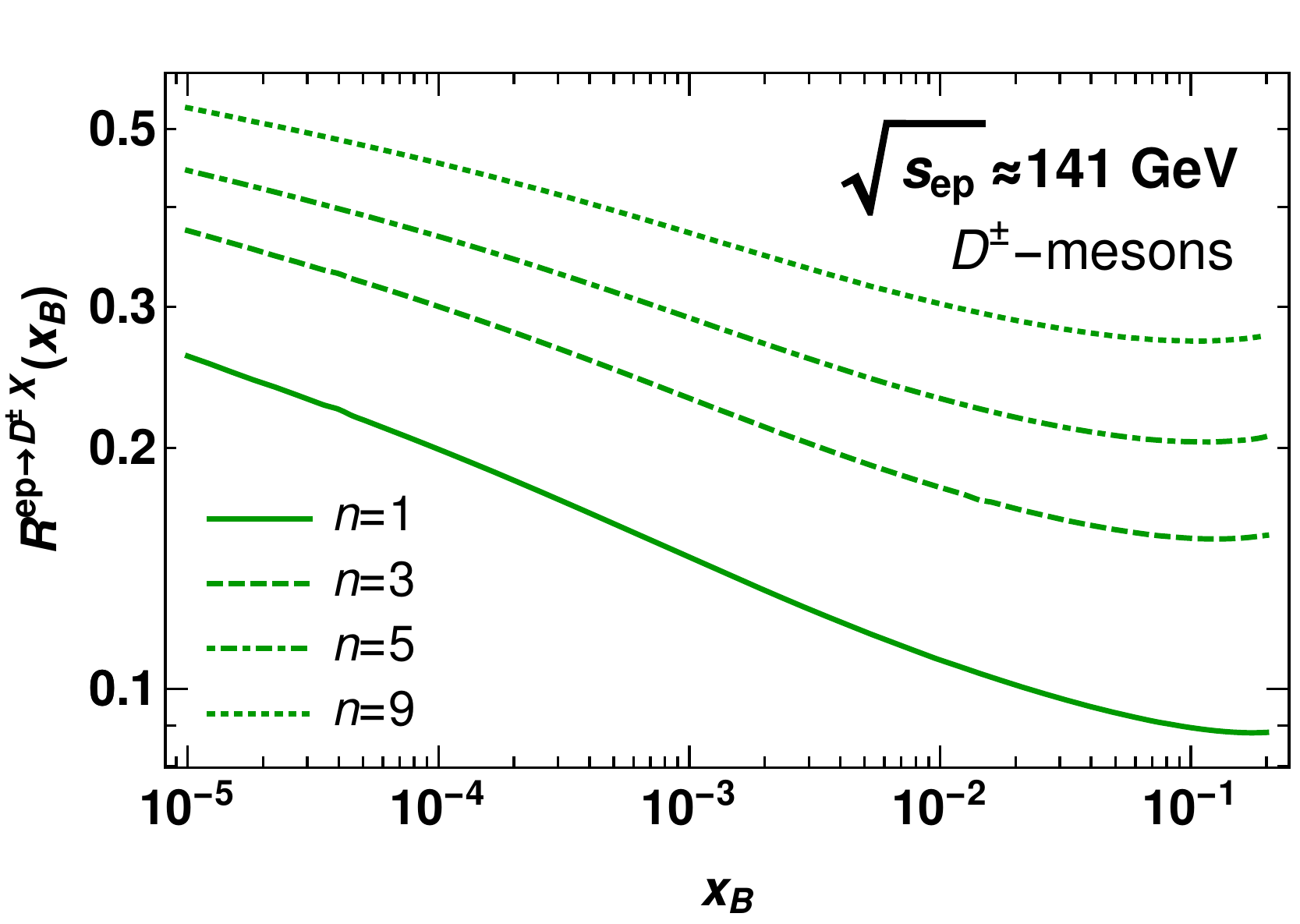}\includegraphics[width=9cm]{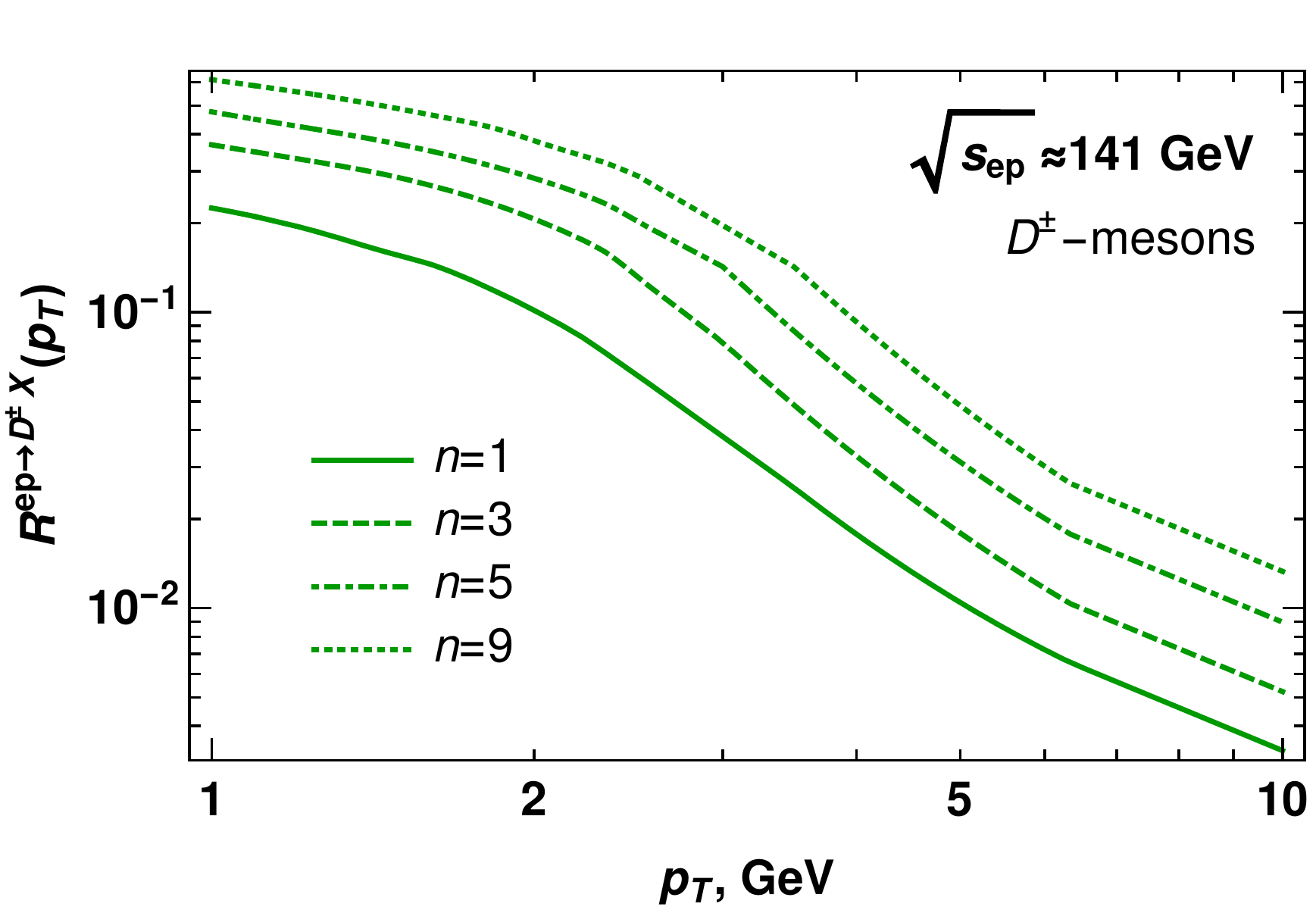}

\includegraphics[width=9cm]{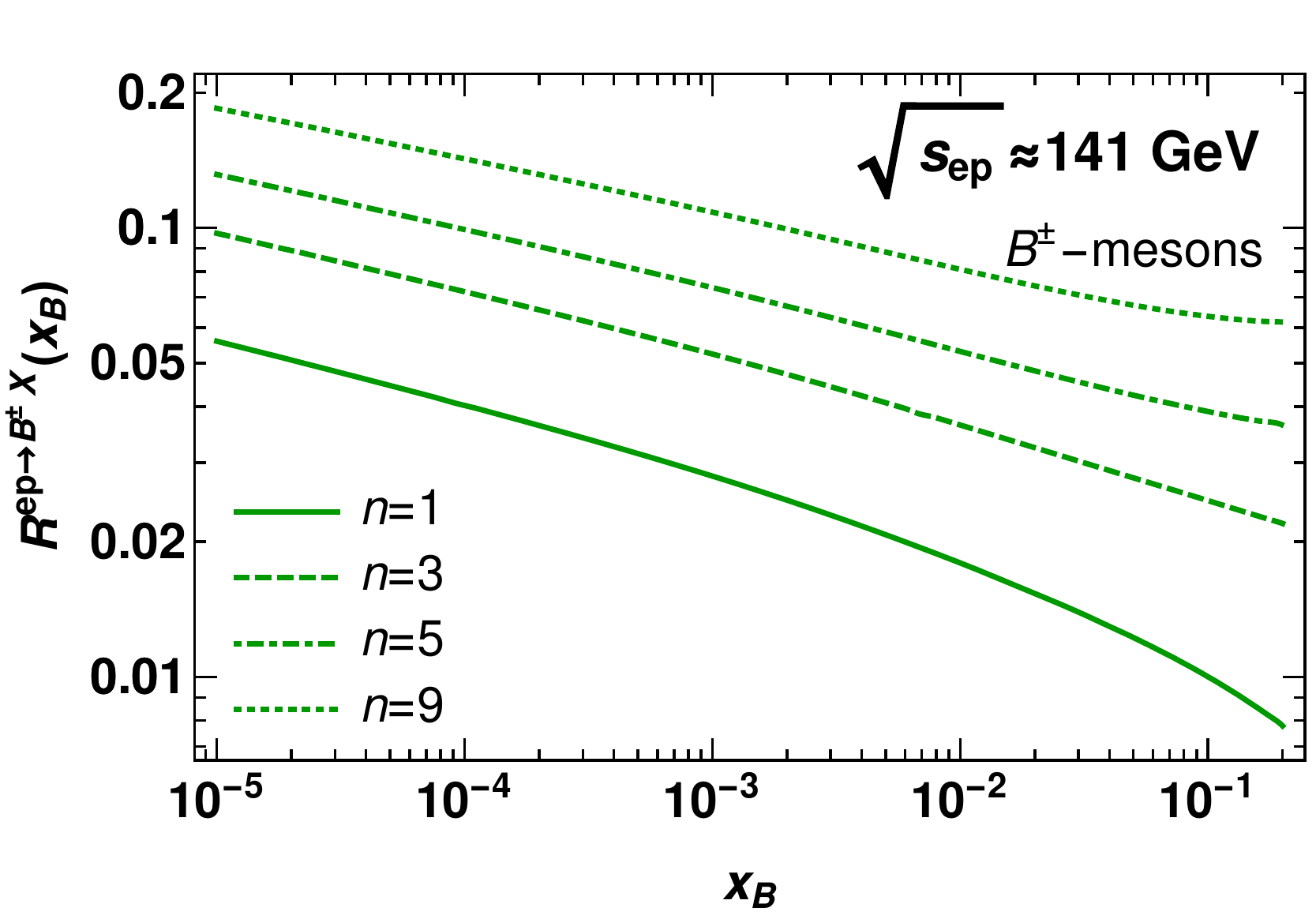}\includegraphics[width=9cm]{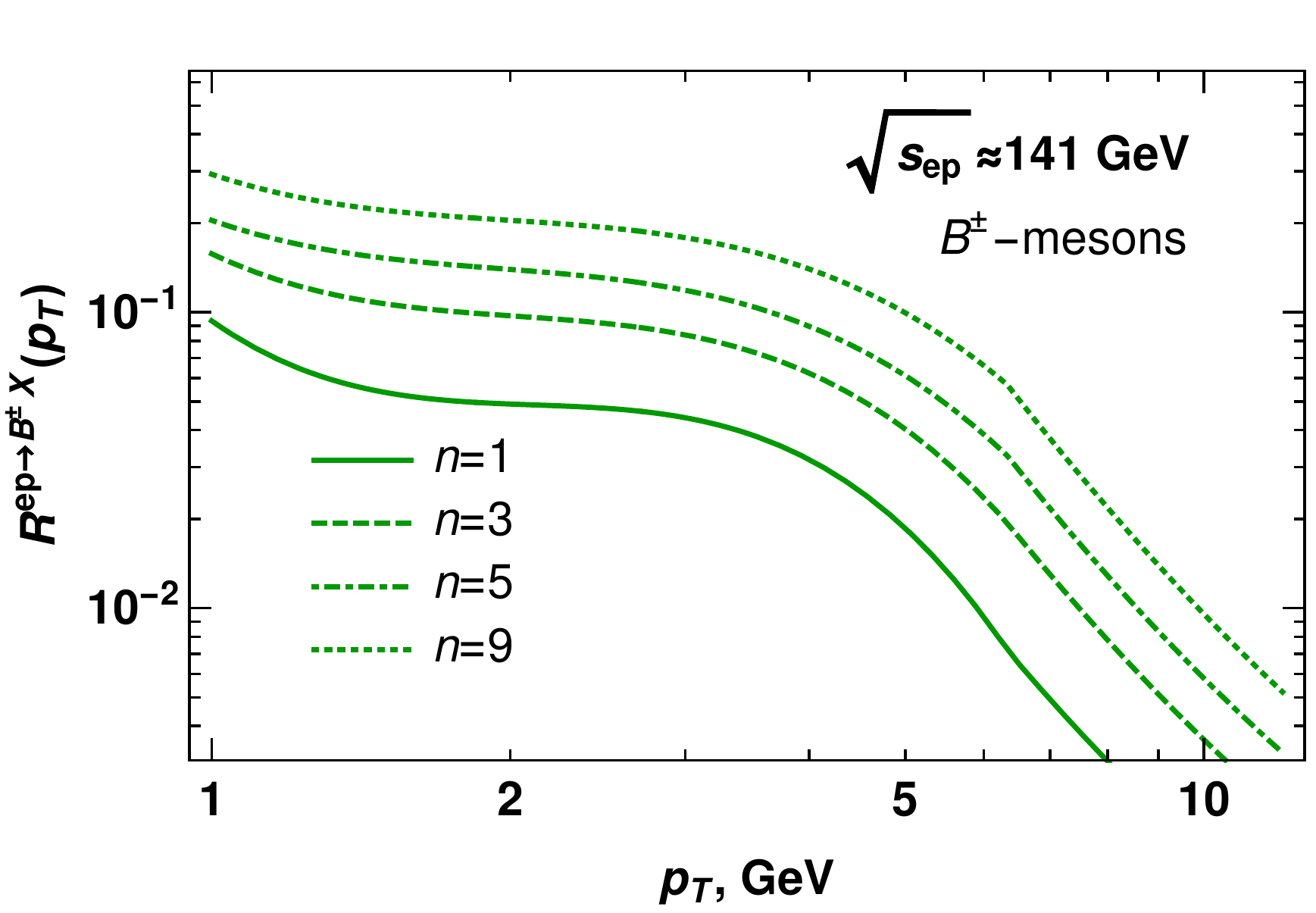}

\caption{\label{fig:Ratio_nDep}The ratio~(\ref{eq:R3}) of the 2-pomeron
to single-pomeron contributions, as a function of $x_{B}$ (left diagram)
and $p_{T}$ (right diagram). The upper row corresponds to $D^{\pm}$-mesons,
the lower row is for $B^{\pm}$-mesons. The variable $n\equiv N_{{\rm ch}}/\left\langle N_{{\rm ch}}\right\rangle $
is the relative enhancement of multiplicity.}
\label{Diags_DMesons-2-1-1-1} 
\end{figure}

\begin{figure}
\includegraphics[width=9cm]{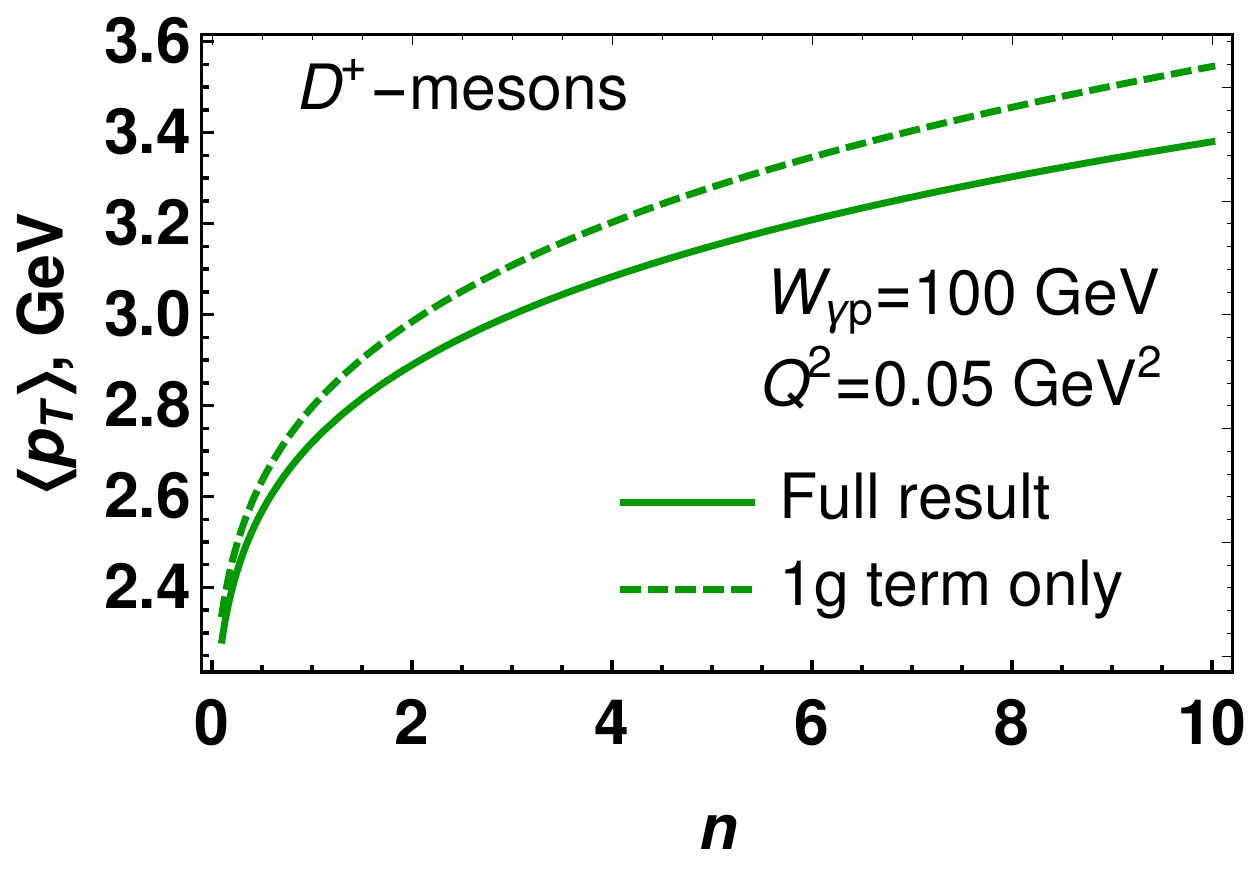}\includegraphics[width=9cm]{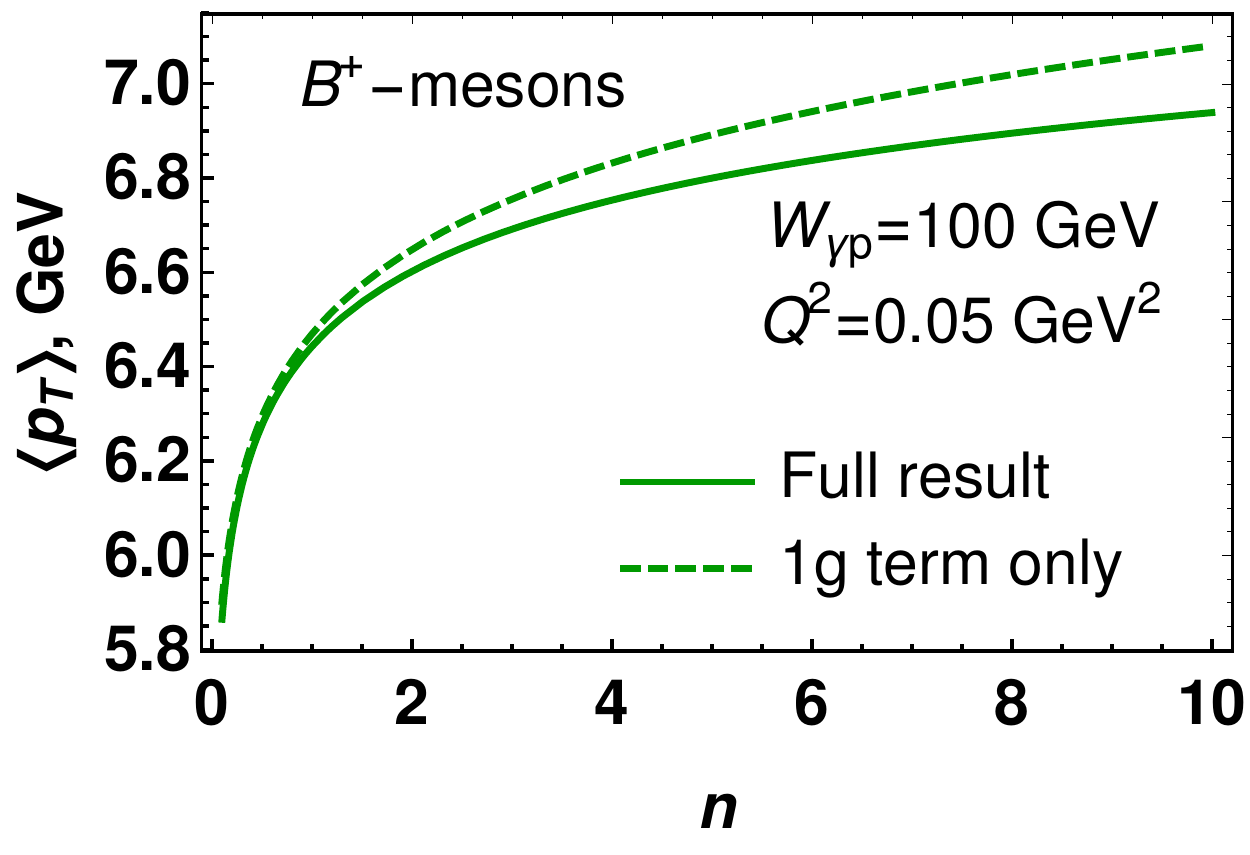}

\includegraphics[width=9cm]{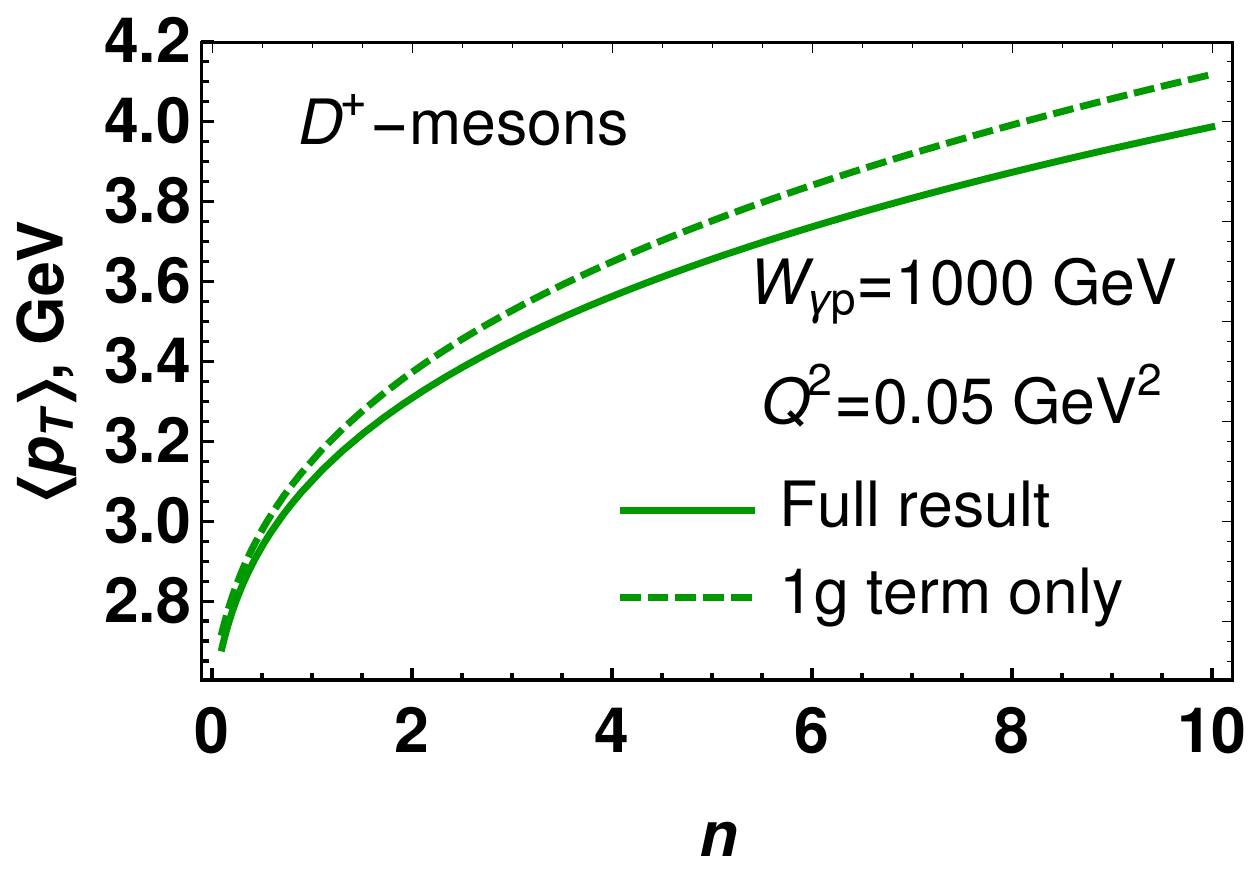}\includegraphics[width=9cm]{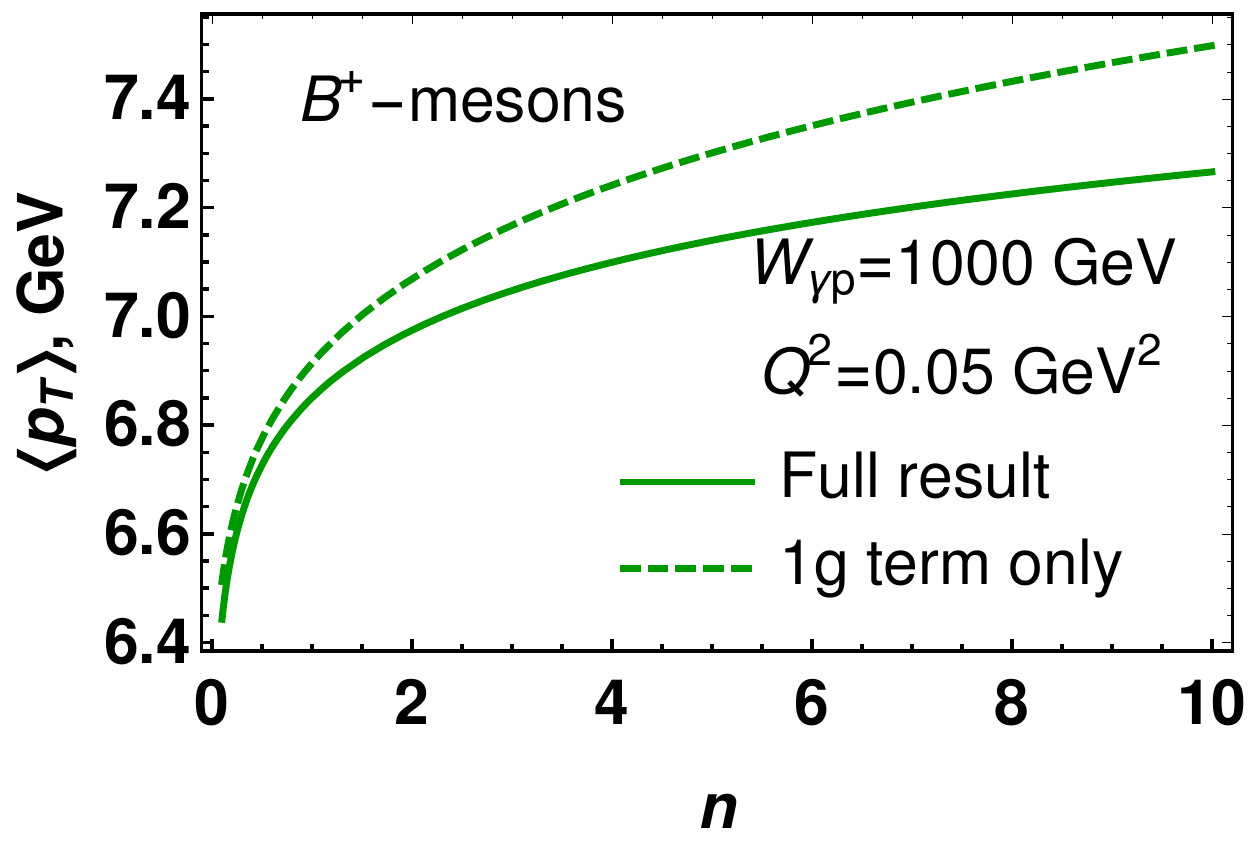}

\caption{\label{fig:nDependencepT} The multiplicity dependence of average
$\langle p_{T}\rangle$ of produced $D$-mesons (left column) and
$B$-mesons (right column). The upper row corresponds to the invariant
photon energy $W_{\gamma p}\approx100$ GeV (EIC kinematics), whereas
the lower row corresponds to higher energy $W_{\gamma p}\approx1000$
GeV, achievable at both LHeC and FCC-he. The dashed curve with label ``Leading
twist'' in all plots, stands for the leading twist (single pomeron) contribution. }
\end{figure}

As we can see from the Figure~\ref{fig:Ratio_nDep}, the theoretical
estimates suggest that in high-multiplicity events the role of the
multipomeron contributions increases. Numerically, in EIC kinematics
this contribution becomes pronounced at $n\gtrsim5$ for $D$-mesons,
although still remains relatively small for $B$-mesons. This difference
in the size of multipomeron terms suggests that we can study experimentally
the multiplicity dependence of the ratio of $D$- and $B$-meson cross-section
in order to estimate unambiguously the role of the two-pomeron contribution
in $D$-meson production. In order to avoid the effects related to the
$x_{B}$-dependence, we suggest to study the double ratio of cross-sections
\begin{equation}
R^{D/B}\left(x_{B},\,n\right)=\frac{d\sigma^{D^{+}}(x_{B},\,n)/d\sigma^{D^{+}}(x_{B},\,n=1)}{d\sigma^{B^{+}}(x_{B},\,n)/d\sigma^{B^{+}}(x_{B},\,1)}.\label{eq:RDB}
\end{equation}
This ratio equals one in the heavy quark mass limit, yet for finite
values of $n$ deviates from this value due to more pronounced higher
twist corrections for $D$-meson (numerator of~(\ref{eq:RDB})).
In the left panel of Figure~\ref{fig:nDependencepT} we 
show the dependence of the ratio~(\ref{eq:RDB}) on $n$. The dependence
on $n$ exists even for the leading twist, due to higher twist corrections,
but becomes more pronounced when the multipomeron contributions are
taken into account. The growth of the ratio as a function of $n$
agrees with the elevated contribution of multipomeron mechanisms in the large-$n$
kinematics. However, we expect that for asymptotically large values
of $n$ the ratio should saturate, because the multipomeron contributions
will also become important in the denominator. In the right panel of the
same Figure~\ref{fig:nDependencepT}, we show a similar self-normalized
double ratio~(\ref{eq:RDB}), in which we replaced $B$-mesons with
non-prompt $D$-mesons. Since the latter mechanism is dominated by
$b$-meson decays, we can see that qualitatively the ratio has a
similar dependence on $n$. Comparison of the left and right panels
of Figure~\ref{fig:nDependencepT} clearly illustrates that the
enhancement of the ratio~(\ref{eq:RDB}) is not related to differences
of the $D$- and $B$-meson fragmentation functions. We expect that
non-prompt charmonia should demonstrate a similar behavior.

\begin{figure}
\includegraphics[width=9cm]{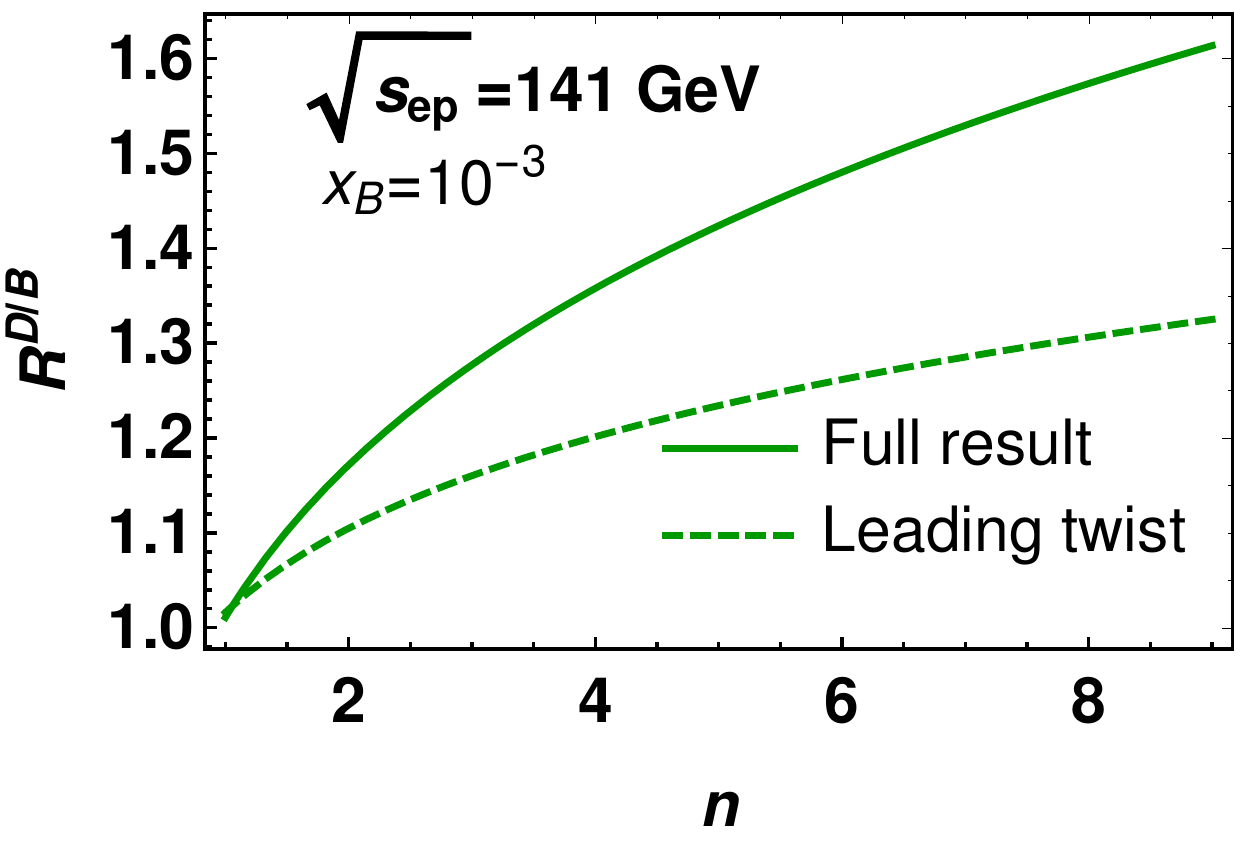}\includegraphics[width=9cm]{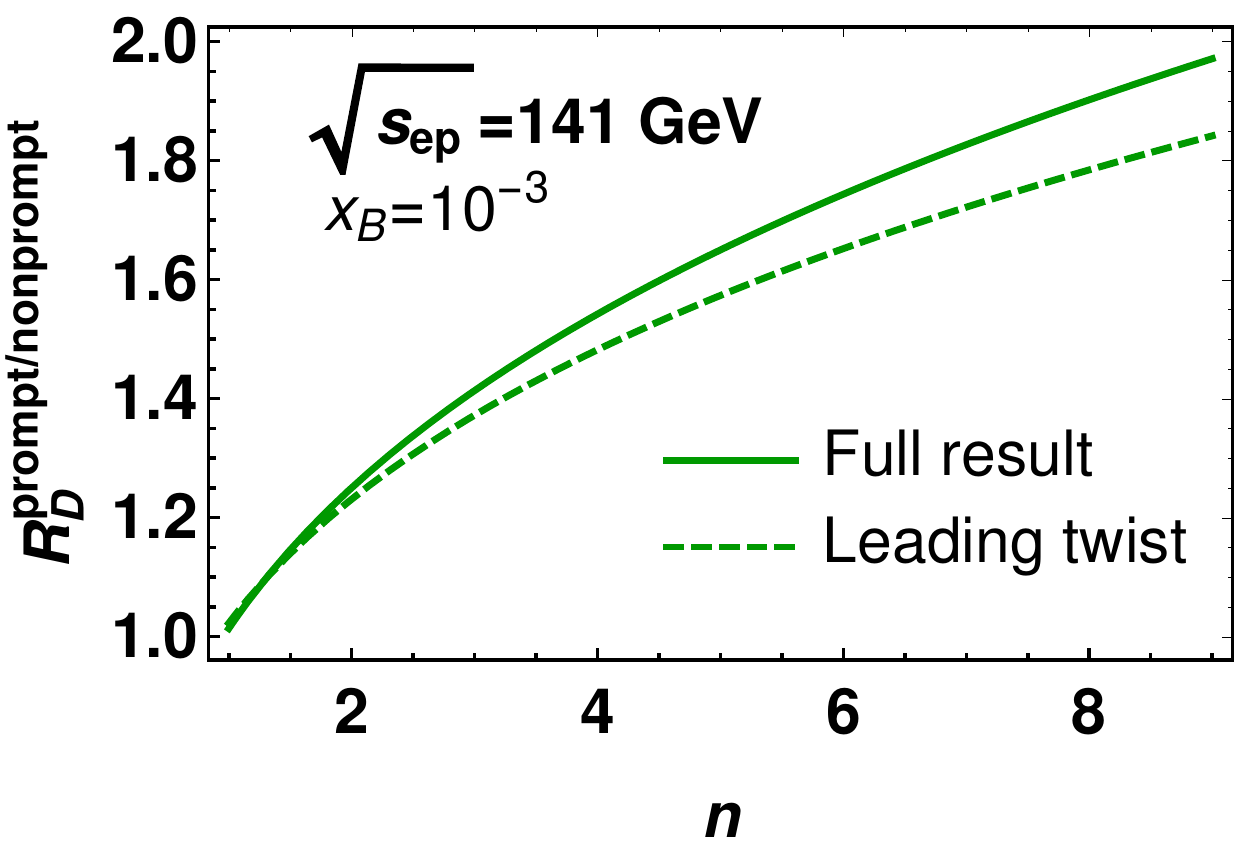}

\caption{\label{fig:nDependence}Left plot: the self-normalized ratio of $D^{+}$-
and $B^{+}$-meson cross-sections, as defined in~(\ref{eq:RDB}).
Right plot: Self-normalized ratio of the prompt and non-prompt $D$-meson
cross-sections, defined similar to~(\ref{eq:RDB}), but with $B$-mesons
replaced by non-prompt $D$-meson cross-section in the denominator. The
dashed curve, with label ``Leading twist'', stands for the leading
twist (single pomeron) contribution. }
\end{figure}

Another observable which might be easily measured experimentally is
the dependence of the average momentum $\langle p_{T}\rangle$ of heavy
mesons on the multiplicity $n$. This observable has been extensively
studied in the context of $pp$ collisions. In Figure~\ref{fig:nDependencepT}
we show the dependence of $\langle p_{T}\rangle$ on $n$, for
electroproduction of both $D$- and $B$-mesons. Since multipomeron contributions
are suppressed at large momenta $p_{T}$, we can see that their inclusion
decreases the average $\langle p_{T}\rangle$, compared to what is expected
from single-pomeron. Although the expected effect is not very large,
we believe that it might be seen in experimental data, since $\langle p_{T}\rangle$
might be measured with very good precision.

To summarize, we believe that the multiplicity dependence might reveal
information about the contribution of the multipomeron mechanisms.
However, in EIC kinematics we do not expect drastic enhancement of
the multiplicity dependence, as  was observed in $pp$ collisions. This
happens because in general multipomeron contributions are small at
EIC energies. The situation might be different in the kinematics of future
accelerators like LHeC and FCC-he, where the role of the multipomeron
contributions is more pronounced. In our analysis we took into account
only the first multipomeron correction, namely the production on two
pomerons. We could see that its relative contribution is small in
EIC kinematics, in agreement with general expectations based on twist
counting, and for this reason we do not consider the orrections of even
higher order. However, at very small values of $x_{B}$ (significantly
smaller than $10^{-7}$) we approach the deeply saturated regime,
where the expectations based on twist expansion are not reliable,
and thus the inclusion of all higher twist might be required.

The mechanism of multiplicity generation suggested in this section
introduces dependence on the multiplicity of soft produced particles,
and is quite different from other approaches, such as the percolation
approach~\cite{PER} or the modification of the slope of the elastic
amplitude~\cite{Kopeliovich:2013yfa}, suggested earlier in the context
of $pp$ studies. We expect that the experimental confirmation of the predicted
multiplicity dependence could help to understand better the mechanisms
of multiplicity enhancement in high energy collisions.

\section{Conclusions}

\label{sec:Conclusions}In this paper we analyzed the mechanisms of
open-heavy flavor meson electroproduction. Motivated by
earlier findings in $pp$ collisions, we also analyzed the relative
contribution of the first subleading multipomeron correction. We found
that for electroproduction this correction is relatively small for
EIC kinematics, although it grows with energy and becomes relevant for
charm production at LHeC and FCC-he, especially in the small-$p_{T}$
kinematics. This correction is less important for $B$-mesons
and non-prompt charmonia production, and does not exceed ten per cent
even at LHeC and FCC-he. The dependence of the correction on $p_{T}$
agrees with general expectations based on large-$p_{T}$ and heavy
quark mass limit. Our evaluation is largely parameter-free and 
describes very well the data from HERA, as well as provides plausible
predictions for EIC, LHeC and FCC-he.

We also analyzed the multiplicity dependence, which might be studied
experimentally in detail at future EIC, LHeC and FCC-he, due to
its outstanding luminosity. The high-multiplicity events present special
interest for theoretical studies, because they allow to get better
understanding of the production mechanisms at high gluon densities. Since
the probability of rare high-multiplicity events is exponentially
suppressed, for the analysis of their dynamics it is important to study
properly the designed variables. We analyzed in detail the dependence
on multiplicity for the average momentum of heavy meson $\langle p_{T}\rangle$
and the double ratio defined in~(\ref{eq:RDB}). The first variable
is easier to measure, although it is less sensitive to higher twist effects,
due to the smallness of subleading contributions. The double ratio~(\ref{eq:RDB})
is more interesting, because its deviations from unity allow to quantify
directly the size of the higher twist corrections, including multipomeron
contributions. Due to the smallness of multipomeron contributions, we
do not expect a significant relative enhancement of the cross-sections
at large multiplicity in EIC kinematics, and only mild enhancement
in the kinematics of LHeC and FCC-he. This expectation differs significantly
from what was found experimentally in $pp$ collisions at LHC~\cite{Adam:2015ota}.
We expect that the experimental confirmation of these findings could help
to understand better the mechanisms of multiplicity generation in
high energy collisions.

\section*{Acknowldgements}

We thank our colleagues at UTFSM university for encouraging discussions.
This research was partially supported by Proyecto ANID PIA/APOYO AFB180002 (Chile) and Fondecyt (Chile) grant 1180232.
Also, we thank Yuri Ivanov for
technical support of the USM HPC cluster where a part of evaluations
has been done. 

 \end{document}